\documentstyle[aps,rmp,epsf]{revtex}

\def\mydelta{{\mit\Delta}\,}
\def\wig#1{\mathrel{\hbox{\hbox to 0pt{%
          \lower.5ex\hbox{$\sim$}\hss}\raise.4ex\hbox{$#1$}}}}
\def\makespace{\phantom{12345678901234567}}

\def\etal{{\it et~al.\,}}
\def\mo{M$_\odot\,$}
\def\lo{L$_\odot\,$}
\def\mj{M$_{\rm J}\,$}
\def\mp{$\,{m}_{p}$}
\def\rp{R$_{\rm p}\,$}
\def\rj{R$_{\rm J}\ $}

\def\teff{T$_{\rm eff}\,$}
\def\teffs{T$_{\rm eff}$s$\,$}
\def\Dwa{$\,$\uppercase\expandafter{\romannumeral5}$\,$}
\def\mic{$\mu$m$\,$}

\def\undertext#1{$\underline{\smash{\hbox{#1}}}$}
\def\sles{\lower2pt\hbox{$\buildrel {\scriptstyle <}
   \over {\scriptstyle\sim}$}}
\def\sgreat{\lower2pt\hbox{$\buildrel {\scriptstyle >}
   \over {\scriptstyle\sim}$}}
\def\sharpnull#1{}

\begin{document}

\title{The Theory of Brown Dwarfs and Extrasolar Giant Planets}
\author{Adam Burrows}
\address{Department of Astronomy and Steward Observatory\\ The University of Arizona\\ Tucson, Arizona, USA 85721
\\ e-mail: aburrows@as.arizona.edu}
\author{W.B. Hubbard, J.I. Lunine}
\address{Lunar and Planetary Laboratory\\ The University of Arizona\\ Tucson, Arizona, USA 85721 
\\ e-mail: hubbard,jlunine@lpl.arizona.edu}
\author{and}
\author{James Liebert}
\address{Department of Astronomy and Steward Observatory\\ The University of Arizona\\ Tucson, Arizona, USA 85721 
\\ e-mail: liebert@as.arizona.edu}
\maketitle

\setcounter{page}{1}
\begin{abstract}

Straddling the traditional realms of the planets and the stars,
objects below the edge of the main sequence have such unique properties,
and are being discovered in such quantities, that one can rightly claim that a new field       
at the interface of planetary science and astronomy is being born.
In this review, we explore the essential elements of the theory 
of brown dwarfs and giant planets, as well as of the new spectroscopic classes L and T. 
To this end, we describe their evolution, spectra, atmospheric compositions, chemistry, physics,  
and nuclear phases and explain the basic
systematics of substellar-mass objects across three orders
of magnitude in both mass and age and 
a factor of 30 in effective temperature.  
Moreover, we discuss the distinctive features of those 
extrasolar giant planets that are irradiated by a central primary, in particular 
their reflection spectra, albedos, and transits. 
Aspects of the latest theory of Jupiter and Saturn 
are also presented.  Throughout, we highlight the effects of condensates, clouds, molecular 
abundances, and molecular/atomic opacities in brown 
dwarf and giant planet atmospheres and summarize   
the resulting spectral diagnostics.  Where possible,  
the theory is put in its current observational context.

\end{abstract}

\tableofcontents





\begin{quote}
\begin{flushleft}  
{\it Purge and disperse, that I may see and tell\\
Of things invisible to mortal sight.\phantom{12345678}}\\
         Milton - Paradise Lost
\end{flushleft}  
\end{quote}

\begin{quote}
\begin{flushleft}  
{\it These blessed candles of the night.}\\
         Shakespeare - Merchant of Venice, Act I, Sc. 1
\end{flushleft}  
\end{quote}

\begin{quote}
\begin{flushleft}  
{\it We are all in the gutter, but some of\\
us are looking at the stars.\phantom{12345678}}\\
         Oscar Wilde - Lady Windermere's Fan 
\end{flushleft}  
\end{quote}

\section{Introduction: A New Frontier for Stellar and Planetary Science}

The term ``brown dwarf''  was coined in 1975 by Jill Tarter (1975) to describe
substellar-mass objects (SMOs), but astronomers had to wait 20 years before the announcement of the discovery
of the first unimpeachable example, Gliese 229B (Oppenheimer \etal 1995).
That same {\it day}, the first extrasolar giant planet (EGP\footnote{We use
this shorthand for \undertext{E}xtrasolar \undertext{G}iant \undertext{P}lanet (Burrows \etal 1995), but terms such as ``super--jupiter''
are also used.}) was announced (Mayor and Queloz 1995) and it startled the
world by being {\it 100} times closer to its primary than Jupiter is to the Sun.

To date, more than 50 EGPs have been discovered by radial-velocity
techniques around stars with spectral types from M4 to F7 (see Table \ref{beast}).
The ``planets'' themselves have minimum masses ($m_p\sin(i)$) between $\sim$0.25 \mj 
and $\sim$10 \mj (\mj $\equiv$ one Jupiter mass), orbital semi-major axes ($a$)
from $\sim$0.04 A.U. to $\sim$3.8 A.U., and eccentricities as high as $\sim$0.71.  Such unanticipated variety
and breadth has excited observers, theorists, and the public at large.

In parallel with this avalanche of planet discoveries are the multitude of recent brown dwarf discoveries
in the field, in young clusters, and in binaries.  More than 150 brown dwarfs are now known.  In fact,
estimates suggest that the number density of brown dwarfs 
in the solar neighborhood is comparable to that of M dwarfs and that in most 
environments the mass function (IMF) is still rising at the 
main sequence edge (Reid \etal 1999). 
Though to date $\sim$17 brown dwarfs have been discovered in binaries
(Zuckerman and Becklin 1992; Rebolo \etal 1998; Basri and Martin 1999; Martin \etal 1999;
Burgasser \etal 2000; Lowrance \etal 1999; Reid \etal 2000b), most seem to inhabit the field.
The growing library of brown dwarf spectra has led to the introduction
of two entirely new ``stellar'' spectroscopic classes: the ``L'' dwarfs (Martin \etal 1999; Kirkpatrick \etal 1999,2000;
Fan \etal 2000; Delfosse \etal 1997; Tinney \etal 1998)
and the ``T'' dwarfs (Burgasser \etal 1999).
Collectively, these new discoveries 
constitute a true renaissance in stellar and planetary astronomy that is becoming the
research focus of a rapidly increasing fraction of the astronomical community. 

This review should be viewed as a modern supplement to the earlier review by Burrows and Liebert (1993).
The latter was written at a time when all brown dwarfs were mere ``candidates.'' Moreover, it was written 
before the modern radial velocity data had transformed the study of planetary systems.  In Burrows and Liebert (1993),
the emphasis was on the relationship between the rudiments of theory (the equation of state, simple
boundary chemistry, the basics of ``stellar'' evolution) and the generic features of brown dwarfs (their radii,
luminosities, effective temperatures).  As such, there was little mention of the composition
and chemistry of the atmospheres, EGPs, optical and infrared spectra and 
colors, and the host of other topics that are now the subject's
modern concerns.  All aspects of SMO research are changing so rapidly that it would be unwise to attempt
to survey in detail all the myriad subtopics that have emerged.  Therefore, we have chosen to concentrate
in this review on a subset of representative topics that, not coincidently, are aligned with 
our own current research interests.  In particular, we will not review in detail the spectra and models of M dwarfs,
very-low-mass (VLM) stars, subdwarfs, low-metallicity stars, or young pre-main-sequence stars.  
Excellent contributions on these topics can be found in D'Antona (1987), 
Allard and Hauschildt (1995), Baraffe \etal (1995), Allard \etal (1997), 
Chabrier and Baraffe (1997,2000), and Chabrier \etal (2000a) and the reader is encouraged to supplement 
our review with these papers.  Furthermore, we will not cover the many theories of planet or star formation.
Here, we cover various major theoretical topics 
concerning SMOs (brown dwarfs and EGPs) that have arisen during the last five years.   These include,
but are not limited to, brown dwarf/EGP evolution, colors, and spectra, the molecular
consistents of their atmospheres, T and L dwarf characteristics,  
reflection spectra, albedos, and transits, recent models of the Jovian planets, the role
and character of the dominant molecular opacities, and heavy-element depletion, rainout,
and clouds.  With such a full plate, we apologize in advance to both reader 
and colleague for omitting a favorite topic and beg their indulgence as we embark upon
our survey of one of the hottest current topics in astrophysics.

\begin{quote}
\begin{flushleft} 
{\it What's in a name? That which we call a rose\\
     By any other name would smell as sweet.\phantom{1234}}\\
         Shakespeare - Romeo and Juliet, Act II, Sc. 2
\end{flushleft}  
\end{quote}

\section{An Overview of Basic SMO Theory}  
\label{over}

Whatever the mass ($M$) or origin of an extrasolar giant planet or brown dwarf, the
same physics, chemistry, and compositions obtain for both.  This fact is at the root
of our ability to encompass both classes with a single theory.  
SMO atmospheres are predominantly molecular, their 
cores are convective metallic hydrogen/helium mixtures
for most combinations of mass and age,  
their emissions are mostly in the near infrared,
and, though deuterium and lithium burning can play interesting roles,
thermonuclear processes don't dominate their evolution (\S\ref{burn}).   After birth,
unless their luminosities ($L$) are stabilized by stellar insolation, they cool off inexorably
like dying embers plucked from a fire.  Figure \ref{lum.ps} depicts the evolution of the luminosity
of isolated solar-metallicity objects from 0.3 \mj ($\sim$ Saturn) to 0.2 \mo  (\mo $\equiv$ 1.0 solar mass $\equiv$ 1047 \mj).
Distinguished by color are objects with masses equal to or
below 13 \mj (red), objects above 13 \mj and below the main sequence edge (green), and stars above the main sequence edge (blue).
These color categories merely guide the eye and clarify what would otherwise be a figure
difficult to parse. The 13 \mj cutoff is near the deuterium-burning limit, but otherwise should not be viewed
as being endowed with any overarching significance.  In particular, objects below 13 \mj that are
born in the interstellar medium in a manner similar to the processes by which stars arise should be referred
to as brown dwarfs.  Objects that are born in protostellar disks by processes that may differ
from those that lead to stars (perhaps after nucleating around a terrestrial superplanet) should
be referred to as planets.  Though theoretical prejudice suggests that such planets may not
achieve masses near 13 \mj, if they do so they are most sensibly called planets.  Hence,
even if the mass distribution functions of EGPs and brown dwarfs overlap (which they no doubt do), a distinction
based upon mode of formation, and not mass, has compelling physical merit, despite the fact that
we can't currently identify the history or origin of any given SMO.  There
is no compelling need to type, define, or classify an SMO immediately upon its discovery. It can
be studied by all available means in the absence of a definitive box into which to put it.  Indeed,
it is through the study and characterization of SMOs and with the help of theory
that natural discriminants of ``planets'' will emerge, be they orbital, spectroscopic, rotational,
or compositional.  Despite the fact that the press and populace are more fixated
on ``planets'' than brown dwarfs,  and some are overly interested in definitions,
at this early stage in the study of SMOs we recommend that the active researcher 
maintain a healthy flexibility vis \`a vis nomenclature.  

Figure \ref{lum.ps} summarizes many aspects of the evolution of solar-metallicity SMOs with masses 
from 0.3 \mj to 0.2 \mo.  The bifurcation between stars and brown dwarfs manifests itself
only at late times ($\sgreat 10^9$ yrs).  At lower metallicities, the brown dwarf/star luminosity gap 
widens earlier and is more pronounced (\S\ref{edge}).  After $10^{8.3}$ to $10^{9.5}$ years, stars stabilize at a 
luminosity for which the power derived from thermonuclear burning in the core
compensates for the photon luminosity (losses) from the surface.  Brown dwarfs are those objects that
do not burn light hydrogen at a rate sufficient to achieve this balance, though the more massive
among them ($\sgreat 0.065$ \mo) do burn light hydrogen for a time.  Figure \ref{tc.ps} depicts
the evolution of the central temperature ($T_{\rm c}$)  (for the same 
set of masses portrayed in Fig. \ref{lum.ps}).  Due to the negative specific
heat effect for an ideal-gas object in hydrostatic equilibrium, radiative losses from the surface lead
to an increase in the core temperatures and densities.  This in turn drives the thermonuclear
power up, while the total luminosity decreases.  If the mass is sufficiently high, the thermonuclear
power will eventually equal the total luminosity, the hallmark of a star.  However, if the mass is too low, $T_{\rm c}$
will not achieve a value sufficient for the thermonuclear power to balance 
surface losses before the core becomes electron-degenerate,
after which time it will cool without compensation by compressional heating.  Figure \ref{tc.ps} shows the rise and fall of 
$T_{\rm c}$ for brown dwarfs.  The peak $T_{\rm c}$ is approximately $2\times 10^6{\rm K}$ ({\rm M}/0.05\mo)$^{4/3}$.
As Fig. \ref{tc.ps} shows, at the hydrogen burning main sequence edge mass (HBMM), the temperature actually decreases 
before stabilizing.  This is possible because the core density continues to increase after the peak $T_{\rm c}$ is achieved, 
compensating in the thermonuclear rate for the slight decrease in $T_{\rm c}$.  Note that for solar metallicity $T_{\rm c}$ at 
the HBMM is as low as $\sim 3\times 10^6$ K, the edge mass is near 0.075\mo (depending upon the
atmospheric opacities, in particular due to silicate grain opacity), the edge \teff is $\sim$1600-1750 K,
and the edge luminosity is near $6\times 10^{-5}$\lo (\S\ref{edge}).   

Figure \ref{rad.ps} portrays the evolution of the radius ($R$) of a red dwarf, brown dwarf, or EGP for the same
set of masses employed in Fig. \ref{lum.ps} (with the same color scheme).  The early plateaus near two to six
times Jupiter's radius coincide with deuterium burning, which roughly stabilizes \teff, $L$, and 
$R$ for from a few$\times 10^6$ years to $10^8$ years, depending upon the mass.  The age when deuterium burning has consumed
50\% of an object's stores of deuterium is indicated in Fig. \ref{lum.ps} by the golden dots.   Deuterium
burning is responsible for the early bumps in Fig. \ref{lum.ps} at high $L$, but, given the likely deuterium
abundance ($\sles 3\times 10^{-5}$), it never leads to a deuterium-burning main sequence.
Nevertheless, deuterium burning will have a measurable effect on the mass-function/luminosity-function
mapping in young clusters and might result in a measurable depletion of deuterium in the atmospheres
of older objects more massive than $\sim$13 \mj (Chabrier \etal 2000b).  
Deuterium burning is also responsible for the increased density of lines in
the lower left hand corner of Fig. \ref{tc.ps} near $T_{\rm c} = 5\times 10^5$ K.

Figure \ref{rad.ps} demonstrates the evolution of radius with age and its non-monotonic dependence  
on mass.  Crudely, at early times the radius is a monotonically increasing function of mass  
and, for a given mass, the radius is always a decreasing function of age.
However, at later times, the dependence of mass upon radius inverts, with the less massive
SMOs having the larger radii.  On the VLM main sequence, the radius of a star increases with mass ($R \propto M^{0.6}$).
For a cold SMO, the peak radius is at a mass of $\sim$4 \mj (Zapolsky and Salpeter 1969; Hubbard 1977).
Importantly, as Fig. \ref{rad.ps} shows, for a broad range of masses from 0.3 \mj to 70 \mj, the older radii are independent of mass to
within about 30\%.  This fact results in a major simplification in our mental image of the class
and is a consequence of the competition in the equation of state between
Coulomb and electron degeneracy effects.  The former would set a fixed density and interparticle distance scale 
(at $\sim$ 1 \AA ) which would lead to the relation: $R \propto M^{1/3}$.  The latter would of its own lead  
to the classic relationship for a low-mass white dwarf ($R \propto M^{-1/3}$).  The competition
between these two trends renders the radius constant over roughly two orders of magnitude
in mass near the radius of Jupiter, a feature of a polytrope of index 1.0.  

As Fig. \ref{lum.ps} shows, the late-time cooling phases for SMOs follow approximate power laws. Indeed, basic
solar-metallicity power-law relations which characterize older SMOs have been  
obtained (Burrows and Liebert 1993; Marley \etal 1996): 

\begin{equation}
L \sim 4\times 10^{-5}\ {\rm L_{\odot}} \Bigl(\frac{10^9 {\rm yr}}{t}\Bigr)^{1.3}\Bigl(\frac{M}{0.05 {\rm M}_{\odot}}\Bigr)^{2.64}
\Bigl(\frac{\kappa_{\rm R}}{10^{-2} {\rm cm^2 gm^{-1}}}\Bigr)^{0.35} , 
\label{powerl}
\end{equation}
\begin{equation}
{\rm T_{eff}} \sim 1550\ {\rm K} \Bigl(\frac{10^9 {\rm yr}}{t}\Bigr)^{0.32}\Bigl(\frac{M}{0.05 {\rm M}_{\odot}}\Bigr)^{0.83}
\Bigl(\frac{\kappa_{\rm R}}{10^{-2} {\rm cm^2 gm^{-1}}}\Bigr)^{0.088} ,
\label{powert}
\end{equation}
\begin{equation}
M \sim 35\ {\rm M_J}  \Bigl(\frac{g}{10^5}\Bigr)^{0.64} \Bigl(\frac{{\rm T_{eff}}}{1000 {\rm K}}\Bigr)^{0.23}  ,
\end{equation}
\begin{equation}
t \sim 1.0\ {\rm Gyr} \Bigl(\frac{g}{10^5}\Bigr)^{1.7} \Bigl(\frac{1000 {\rm K}}{{\rm T_{eff}}}\Bigr)^{2.8}  ,
\end{equation}
\begin{equation}
R \sim 6.7\times 10^4\ {\rm km} \Bigl(\frac{10^5}{g}\Bigr)^{0.18} \Bigl(\frac{{\rm T_{eff}}}{1000 {\rm K}}\Bigr)^{0.11},
\label{powerr}
\end{equation}
where $g$ is the surface gravity in cgs units and $\kappa_{\rm R}$ is an average atmospheric Rosseland mean opacity.
Different prescriptions for the atmospheric opacity ({\it e.g.}, for silicate clouds) and the equation 
of state, as well as different metallicities, will result in different numbers,
but the basic systematics for the late-time evolution of SMOs is fully encapsulated in the above equations.
Note that if one drops the $\kappa_{\rm R}$ dependence and divides by $\sim$4, 
the luminosity relationship in eq. (\ref{powerl}) can be used 
for zero-metallicity SMOs (Saumon \etal 1994; Burrows \etal 1993).
A better zero-metallicity fit is obtained if one substitutes the indices 1.25 and 2.4 for 1.3 and 2.64, respectively. 
The corresponding zero-metallicity formula for \teff is 
obtained by substituting 1140 K for 1550 K in eq. (\ref{powert}), with the indices 0.31 and 0.77 substituted
for 0.32 and 0.83, respectively.
The analytic formulae in eqs. (\ref{powerl}-\ref{powerr}) allow one to derive 
any quantity from any two other quantities, for a given $\kappa_{\rm R}$
or metallicity (Z). Given a composition, the EGP/brown dwarf continuum
is a {\it two}-parameter family that spans two orders of magnitude in mass, three orders of magnitude
in age, and more than one and a half orders of magnitude in effective temperature from $\sim$80 K to $\sim$3000 K.

A power-law relationship can be derived linking luminosity and mass on the lower main sequence (see Fig. \ref{lum.ps}).
For solar metallicity, Burrows and Liebert (1993) obtain: 
\begin{equation}
L_{\rm star} \sim 10^{-3} {\rm L_\odot} \Bigl(\frac{M}{0.1 {\rm M_{\odot}}}\Bigr)^{2.2}.
\label{above}
\end{equation}
A comparision of eq. (\ref{above}) with eq. (\ref{powerl}) reveals that the dependence of $L$ on $M$ 
is slightly steeper below the main sequence than above.

\section{The Edge of the Main Sequence}  
\label{edge}

The properties of stars at the main sequence edge are a function of helium fraction ($Y_{\alpha} \sim 0.25-0.28$),
metallicity, and the opacity of the clouds of silicate grains that characterize L dwarf atmospheres with \teffs near
1400-2000 K (\S\ref{spectra}).  A large grain opacity (perhaps due to smaller average particle size), 
high helium fractions, and higher metallicity lead to lower edge masses (HBMM),
edge \teffs, and edge luminosities.  The higher helium fraction leads to a more compact object,
with a larger central temperature and density, all else being equal.  
Larger opacities translate into optically thicker atmospheric blankets that do two things:
1) they produce higher central temperatures by steepening the temperature gradient and 2) they decrease the
energy leakage rate (luminosity) from the surface.  The former enhances the thermonuclear rate while the 
latter makes it easier to achieve main-sequence power balance at a lower mass.  Hence, the HBMM at solar metallicity
and $Y_{\alpha} = 0.25$ is $\sim$0.07-0.074 \mo (Hayashi and Nakano 1963; Kumar 1963), with a T$_{\rm eff}$ of 1700-1750 K 
and an $L_{\rm edge}$ of $\sim\!6.0\times 10^{-5}$ \lo, while the HBMM at zero metallicity is $\sim 0.092$\mo,
with a T$_{\rm eff}$ of $\sim$3600 K and an $L_{\rm edge}$ of $\sim 1.3\times 10^{-3}$ \lo (Saumon \etal 1994).
The derivative, ${\partial {\rm M_{edge}}}/{\partial Y_{\alpha}}$, is approximately equal to $-0.1$ \mo.
However, as implied above, uncertainties in silicate grain physics and opacities 
leave us with ambiguities in the HBMM, its \teff, and $L$.  As a consequence, 
the effective temperature at the HBMM might be as low as 
1600 K (Baraffe \etal 1995; Chabrier and Baraffe 2000; Chabrier \etal 2000a).  
However, whatever the role of silicate clouds at the main sequence edge,
a solar-metallicity edge object is an L dwarf. (We will discuss L dwarfs in greater detail
in \S\ref{observed}.)
The edge characteristics and the late boundary of
the L dwarfs both depend upon the persistence and properties of silicate grains.
As an added consequence, the width in \teff space of the L dwarf sequence is probably an increasing
function of metallicity.

If the edge of the main sequence is an L dwarf, then only a fraction of the L dwarfs are substellar.
To illustrate this, let us turn to
the case of the prototypical L dwarf, GD 165B.  With a spectral class of
L4, this object is in the middle of the L scale (Kirkpatrick \etal 1999). 
Since GD 165A is a white dwarf with a \teff of $\sim$11,000 K,
GD 165B is a long-lived object; the white dwarf 
cooling age requires a systemic age of $\sim$ 1.0 Gyr.  
Since the white dwarf mass estimated from the surface gravity is not
particularly high, a nuclear lifetime of at least a few Gyrs is
indicated for the white dwarf progenitor.  
The trigonometric parallax of the
primary allows the determination of GD 165B's 
absolute $J$ magnitude (M$_J$ = 13.31$\pm$0.18)
and it is fainter than that of any M dwarf.
This puts a very strong constraint on its mass:
if it is not a main sequence star, a prolonged
period of nuclear burning is required for it not to have cooled to a
much lower luminosity.  Therefore, GD 165B's mass could
be around 0.07-0.075 \mo (for solar composition).

Reid \etal (1999) addressed the current ambiguity in the fraction
of L dwarfs that are substellar by utilizing observations of
the lithium line (6708 \AA\ in the red spectrum) in the so-called 
``lithium test" (Rebolo, Martin, and Magazzu 1992; \S\ref{burn}).
About one third of the 2MASS L dwarfs show detections of lithium.  These
arguably have masses below the 0.06-0.065 \mo mass limit for the
interior to be hot enough for lithium burning (see Fig. \ref{yl} below).  If lithium is
depleted in these completely convective objects, the mass is above this
value.  This expectation is nicely vindicated for GD 165B by the small upper limit (0.7 \AA)
to the equivalent width of the lithium line measured by Kirkpatrick \etal (1999),
though the proximity of its stellar companion slightly complicates the  
interpretation. 

That the HBMM is a weak function of metallicity was demonstrated analytically by Burrows and Liebert (1993),
who derived an approximate formula: 
\begin{equation}
{\rm M}_{\rm edge} \sim 0.0865 \Bigl(\frac{10^{-2} {\rm cm^2 gm^{-1}}}{\kappa_{\rm R}}\Bigr)^{1/9} {\rm M}_{\odot},
\end{equation}
where $\kappa_{\rm R}$ is an ersatz for metallicity.  Hence, it is not surprising that the HBMM ranges from
0.092 \mo to only 0.07 \mo as Z changes by orders of magnitude.  However, \teff and $L$ at a given mass and age
can vary significantly with Z.  Figure \ref{LM.metal} shows 
$10^{10}$-year $L$--$M$ isochrones as a function of metallicity
for two opacity prescriptions (Alexander and Ferguson 1994; Allard 
and Hauschildt 1995), as calculated by Burrows \etal (1998)
(using the atmospheric boundary conditions provided by Didier Saumon).   
Included is the zero-metallicity isochrone from Burrows \etal (1993).  Red dwarf stars are in the top right and 
brown dwarfs are in the bottom left.  Note that the transition region is a sensitive function of
metallicity, as is the luminosity at a given mass in both the stellar and substellar regimes.   
Note also that $L$ is a decreasing function of metallicity above the HBMM, but an increasing function
of metallicity below it.  Low metallicity means low opacity that allows a brown dwarf to cool more
quickly.  By the same token, low opacity allows one to see more deeply into stabilized stars to higher temperature layers,
making subdwarfs and extreme subdwarfs more luminous. The corresponding T$_{\rm eff}$-$M$ isochrones
demonstrate the same systematics and are provided in Fig. \ref{TM.metal}.

Observers probing the edge of the main sequence do so in magnitude space. It is interesting to note
that at the solar-metallicity edge M$_{\rm V}$, M$_{\rm R}$, and M$_{\rm K}$ are $\sim$19.5, $\sim$18, and $\sim$11.5, respectively,
but at the zero-metallicity edge they are 12.8, 12.0, and 11.1. In $V$ and $R$, the edge varies by 6-7 magnitudes over this  
metallicity range, but, curiously, $K$ is roughly the same.  This is a 
consequence of the importance even when there are no metals of 
collision-induced absorption (CIA) by H$_2$ at 2.2 \mic (\S\ref{moleopac}) and is an interesting fixed
point in VLM/SMO theory.

\begin{quote}
\begin{flushleft}  
{\it Some say the world will end in fire,\\
     Some say in ice,\makespace\qquad\qquad\\
     From what I've tasted of desire,\qquad\qquad\qquad\\
     I hold with those who favor fire.\qquad\qquad\qquad}\\
         Robert Frost - Fire and Ice
\end{flushleft}  
\end{quote}

\section{Deuterium and Lithium Burning}
\label{burn}

Though SMOs are characterized by the fact that they don't generate sufficient
power by thermonuclear processes to balance their surface radiative losses,
they can have thermonuclear phases, however partial or temporary.  Objects more massive than 
$\sim$13 \mj will burn deuterium via the $p +d\rightarrow  \gamma +$$^{3}\!He$
reaction and objects more massive than $\sim$0.06 \mo ($\sim$63 \mj) will burn 
lithium isotopes via the $p +$$^{7}\!Li \rightarrow 2\alpha$ and $p +$$^{6}\!Li \rightarrow \alpha +$$^{3}\!He$
reactions. The terrestrial $^{6}Li/^{7}Li$ ratio is $\sim 0.08$.  The gold and magenta
dots on Fig. \ref{lum.ps} indicate the ages at which 50\% of the deuterium and lithium, respectively,
have burned.  Note that the lithium dots extend into the brown dwarf regime.  This fact is the 
origin of the so-called ``lithium test'' by which the presence of the atomic lithium line at 6708 \AA\
in older objects is used as an index of sub-stellarity (Rebolo, Martin, and Magazzu 1992).  Figures \ref{yd} and \ref{yl} portray the
evolution of the deuterium and lithium abundances for solar-metallicity objects, using the same mass set   
found on Fig. \ref{lum.ps}.  In Figs. \ref{lum.ps} and \ref{yl}, 
the behavior of the lithium fraction between ages of $10^8$ and $10^9$ years
explains the lithium test, but Fig. \ref{teff}, which displays the evolution of \teff for the same model set,
provides a more profound view of this phenomenology.  
In particular, we see from Fig. \ref{teff} that
the edge of the lithium sequence is near \teff $\sim$ 2600 K and the edge of the deuterium sequence
is near \teff $\sim$ 2000 K.  Moreover, these edges are reached after $\sim 2\times 10^8$ years and $\sim 3\times 10^7$ years,
respectively.  The corresponding spectral types are near M6 and L0-2.  The luminosities at 
both edges are near $10^{-3}$ \lo .  These are useful facts that characterize the thermonuclear
features of SMOs.  

The lithium test is less useful now that we have penetrated so unambiguously and often into
the substellar realm.  Furthermore, in the mid- to late-T dwarf regime (bounded below by the 
``water cloud'' class, see \S\ref{chem}), lithium is in molecular
form (mostly LiCl; see \S\ref{chem}) and can not be identified by the presence 
of the 6708 \AA\ atomic line. (However, note that LiCl has a band near $\sim$15.5 \mic .) 
Importantly, chemistry dictates that the strength
of the 6708 \AA\ line should peak near the middle or end of the L dwarf sequence 
(Lodders 1999; Burrows, Marley, and Sharp 2000).

\section{Atmospheric Chemistry and Abundances}
\label{chem}

Central topics of SMO theory are atmospheric chemistry, thermo-chemical databases,
and the molecular abundances of the major atmospheric constituents.  For solar metallicity,
near and above brown dwarf/EGP photospheres
the dominant equilibrium form of carbon is CH$_4$ or CO, that
of oxygen is H$_2$O, and that of nitrogen is either N$_2$ or NH$_3$ (ammonia), depending upon T$_{\rm eff}$ (Fegley and Lodders 1996).
Hydrogen is predominantly in the form
of H$_2$.  Silicates, most metals, TiO, and VO are found at temperatures above 1600-2000 K.
Neutral alkali metals are found at temperatures above $\sim$1000 K.
Clouds of NH$_3$ and H$_2$O can form for T$_{\rm eff}$s below
$\sim$200 K and  $\sim$500 K, respectively.  For a solar-metallicity SMO to achieve an effective temperature of $\sim$200-300 K
within $10^{10}$ years, and thus to form NH$_3$ clouds within a Hubble time, it must have a mass less than $\sim$10-15 \mj.
The corresponding SMO mass below which a H$_2$O cloud can form within a Hubble time is $\sim$30-40 \mj .
Hence, we should expect to discover many brown dwarfs capped with H$_2$O clouds.
Such objects (``water cloud'' dwarfs) would constitute another spectroscopic class after the T dwarfs.
The associated mass/age ranges are easily determined using Fig. \ref{consttemp.ps}, which displays lines of constant \teff
in mass--age space for solar-metallicity models. The hooks in Fig. \ref{consttemp.ps}
are a consequence of deuterium burning.  Such a figure
is very useful for determining many features and trends of the SMO family, not just those related to
condensation and clouds.

Thermochemical data with which to treat the condensation of CH$_4$, NH$_3$, H$_2$O, Fe, and
MgSiO$_3$ can be obtained from various sources, including Eisenberg and Kauzmann (1969), the
Handbook of Chemistry and Physics (1993), Kurucz (1970), and Lange's Handbook of Chemistry
(1979). Most of the other thermodynamic data 
can be obtained from the JANAF tables (Chase
1982; Chase {\it et al.} 1985).  Data on a number of condensates not available
in these tables can be found in Turkdogan (1980) and data on the two
condensates NaAlSi$_3$O$_8$ (high albite) and KAlSi$_3$O$_8$ (sanidine)
are found in Robie and Waldbaum (1968).  Tsuji (1973) is a source
of data for the gas-phase molecules CaH, CrH, MnH, NiH, MnO, NiO, MnS, TiS, TiN,
SiH$_2$, SiH$_3$, together with some carbides of little importance in the brown
dwarf context. With the exception of rubidium (Barin 1995), data for the hydrides, hydroxides, sulfides,
chlorides, and fluorides of the alkalis are found in the JANAF tables.
Chemical equilibrium abundances at a given temperature, pressure, and elemental composition
are obtained numerically by minimizing the total free energy
of a mixture of a suitably large set of species, subject to the
constraint of particle (element) conservation (Burrows and Sharp 1999).
Maintaining and updating complete thermochemical databases is a constant chore, so
for the most up-to-date source the reader is referred to the collected
works of Fegley and Lodders.

One can easily be confused by the multitude of molecules that in principle can form
from the general mix of the elements, but equilibrium calculations for a solar abundance
pattern of the elements (Table \ref{anders}) at the temperatures and pressures encountered in SMO
atmospheres ($\sim$100 K $\sles$ T $\sles$ $\sim$3000 K; $\sim$10$^{-4}$ bar $\sles$ P $\sles$ $\sim$ 10$^3$ bar),
as well as observations of M, L, and T dwarfs, have narrowed the list of relevant compounds considerably.
Figure \ref{elem:abund} depicts in graphical form the fractional elemental number abundances (conveniently
arrayed along the diagonal).  The arrows on this figure point to balloons that contain some of the
major compounds found in SMO atmospheres and associated with these elements, though not all. 
The reader is encouraged to study this information-rich figure.  Most of the major elements (except Rb and Cs)
are represented.   The abundance hierarchy of the elements depicted in both Fig. \ref{elem:abund}
and Table \ref{anders} speaks volumes about which elements/compounds dominate in SMO
atmospheres.  After hydrogen (in the form of H$_2$) and helium, the reason for the importance of O/C/N compounds is
manifest in Fig. \ref{elem:abund}.  Figure \ref{comp.gl229b} portrays the abundance profiles
in a Gl 229B model of the dominant O/C/N molecules (H$_2$O, CO, CH$_4$, N$_2$, and NH$_3$)
and is quite representative for SMO atmospheres. 

Neon is inert, both
chemically and spectroscopically, but Mg and Si are abundant.  The fact that
they are more abundant than Ca and Al means that even if all the Ca/Al is sequestered in compounds
there is excess Mg and Si available to form Mg/Si/O compounds.  Since Ca and Al silicates are generally
more refractory than Mg$_2$SiO$_4$ (forsterite) and MgSiO$_3$ (enstatite), they and their compounds are expected
to settle and ``rain out'' first at high temperatures (1800-2500 K).  Since Mg, Si, and O are in ample supply in a solar mix,
enstatite and forsterite will be expected at lower temperatures, higher up in the atmospheres. Hence, they
may be expected to dominate in most of the silicate clouds ``seen'' in SMO atmospheres, perhaps for
spectral subtypes later than the early Ls.  

The overwhelming presence of hydrogen ensures that the light hydrides (H$_2$O, CH$_4$, NH$_3$, H$_2$S)
as well as the heavier hydrides ({\it e.g.}, FeH, CrH, CaH, MgH) play roles.  At low
metallicities, the latter will be even more common.  In fact, all of these
compounds have been detected in either L or T dwarfs or in Jupiter (H$_2$S).  Since C/O is less than one,
hydrocarbon chemistry is not expected, unless UV-driven non-equilibrium processes, such as are expected
in irradiated EGPs, occur. Hence, carbon is in the form of CO at high temperatures and low pressures and of CH$_4$ at
low temperatures and high pressures.  The pressure dependence of the CO/CH$_4$ ratio is a straightforward
consequence of Le Ch\^atelier's Principle, given the reaction CH$_4$ + H$_2$O $\rightleftharpoons$ CO + 3H$_2$.
(However, Noll, Geballe, and Marley (1997) have shown that the abundance of CO in Gl 229B is 
enhanced above the expected equilibrium value, perhaps due to disequilibrium kinetics driven by the speed
of convective upwelling.)  

Oxygen is predominantly in the form of water and CO, but is available
in sufficient abundance to enable the formation of a variety of oxides, mostly at higher temperatures
({\it e.g.}, Al$_2$O$_3$, TiO, VO, SiO(g)).  TiO and VO are so spectroscopically active that they
are used to define the M dwarf sequence in the optical and near IR 
(see \S\ref{spectra} and Fig. \ref{opac.sum} in \S\ref{spectra}).  Their disappearance near the
M/L border (TiO near 2000-2200 K and VO near 1800 K) helps to announce and define the L dwarfs. 
At lower temperatures, they either condense out directly (VO) or form condensable solids (for M dwarfs:CaTiO$_3$ [perovskite],
for L/T dwarfs: Ti$_3$O$_5$, Ti$_2$O$_3$).

FeH and CrH are seen in late M dwarfs and are present in L dwarfs and M subdwarfs.  Chemistry implies that CrH will persist
to lower temperatures ($\sim$1500 K) than FeH ($\sim$2000-2200 K).  However, the line lists for
both compounds are incomplete and this translates into errors in their partition functions and
free energies.  The upshot might be that FeH in particular should persist to even lower temperatures.
Furthermore, the predominant reservoir of iron below 2300 K should be metallic Fe droplets, which are thermodynamically
favored over FeH at these temperatures.  These droplets should settle (rain out) in the
gravitational field and reside in a cloud near the 2000-2200 K level
(even in Jupiter), depleting the upper SMO atmospheres of iron and iron compounds (Fegley and Lodders 1994; Burrows and Sharp 1999).
Being more refractory than enstatite and forsterite, this cloud
should undergird most of the silicate cloud layers.  Since FeS is the preferred
form of iron below temperatures of $\sim$750 K and S does not form refractory compounds above
such a temperature, if iron did not rain out at depth, we would not see H$_2$S in Jupiter's atmosphere (Fegley and Lodders 1994).
Sulfur would be in the form of FeS. However, in fact we do see abundant H$_2$S
in Jupiter's atmosphere and this implies that iron did indeed
rain out to depth in the atmospheres of the Jovian planets
and is not available to scavenge sulfur at altitude.
Similarly, iron will rain out in L and T dwarfs and H$_2$S is their major sulfur reservoir.

As shown by Burrows and Sharp (1999), Fegley and Lodders (1996), and Lodders (1999), the alkali metals
are less refractory than Ti, V, Ca, Si, Al, Fe, and Mg and survive in abundance as neutral atoms in substellar
atmospheres to temperatures of 1000 K to 1500 K.  This is below the 1600 K to 2500 K temperature
range in which the silicates, iron, the titanates, corundum (Al$_2$O$_3$), and spinel (MgAl$_2$O$_4$), etc. condense and rain out.
The rainout of refractory elements such as silicon and aluminum ensures that Na and K are not sequestered in the feldspars
high albite (NaAlSi$_3$O$_8$) and sanidine (KAlSi$_3$O$_8$) at temperatures
at and below 1400 K, but are in their elemental form down to $\sim$1000 K.
Hence, in the depleted atmospheres of the cool T dwarfs and late L dwarfs, alkali metals quite naturally
come into their own.  Figures \ref{rainout} and \ref{norain} demonstrate the role of rainout by
depicting the profiles of the relative abundances of the main reservoirs
of the alkali metals, with and without rainout as crudely defined in Burrows and Sharp (1999).
As is clear from a comparison of these two figures, rainout and depletion of heavy metals can result
in a significant enhancement in the abundances at altitude (lower temperatures) of the neutral alkali metal atoms,
in particular sodium and potassium.
Figures \ref{rainout} and \ref{norain} indicate that the alkali metals
Na, K, Li, and Cs (and presumably Rb) form chlorides, hydroxides, and sulfides below temperatures of 1500-1700 K.
The \teffs and gravities at which this happens depends upon the specific atmospheric temperature/pressure profiles, the
character and efficiency of rainout, and the basic chemistry. This implies that the strengths of the alkali atomic lines, so
important in L and T dwarfs (see \S\ref{shape}, \S\ref{spectra}, and \S\ref{determ}), eventually diminish, but that they diminish in
a specific order.  Moreover, the coldest T dwarfs and ``water cloud'' dwarfs at even lower \teffs should have weak alkali lines. 

Figure \ref{profile:cloud} depicts various condensation or chemical equilibrium/transition lines for the alkali chlorides,
enstatite/forsterite, some Ti compounds, and the major O/C/N compounds.  Superposed are atmospheric temperature/pressure profiles
from Burrows \etal (1997) for 1-Gyr brown dwarfs and red dwarfs and for Jupiter.  The intersection point of a
given condensation curve with an SMO's T/P profile indicates the likely position of the corresponding cloud base.
As expected, NH$_3$ and H$_2$O clouds form at low temperatures and pressures and silicate clouds form at high temperatures
and pressures. Since the T/P profile of a low-mass, old SMO is associated with a low-entropy
core adiabat, clouds form at higher pressures in low-mass and/or older brown dwarfs/EGPs.  In fact,
in keeping with these expected systematics, NH$_3$ clouds exist at higher pressures in Saturn than in
Jupiter.

\begin{quote}
\begin{flushleft}
{\it When clouds appear like rocks and towers,\\
     The earth's refreshed by frequent showers.}\\
         Old Weather Rhyme
\end{flushleft}
\end{quote}

\begin{quote}
\begin{flushleft}
{\it Far clouds of feathery gold,\\
     Shaded with deepest purple gleam\\
     Like islands on a dark blue sea.}\\
         Percy Bysshe Shelley - Queen Mab
\end{flushleft}
\end{quote}

\begin{quote}
\begin{flushleft} 
{\it Into each life some rain must fall,\\
     Some days must be dark and dreary.}\\
        Henry Wadsworth Longfellow - The Rainy Day
\end{flushleft}
\end{quote}

\subsection{Cloud Models}
\label{clouds}

The presence of condensed species can radically
alter the gas phase composition.  The more refractory
condensates whose condensation points lie well below the photosphere still
play an important role in depleting the observed atmosphere of a number of
abundant elements, {\it e.g.},  Al, Si, Mg, Ca, and Fe.  The almost complete absence
of spectral signatures of metal oxides (such as
TiO and VO) in Gliese 229B and the other T dwarfs is in keeping with theoretical expectations (Burrows and Sharp 1999; Fegley and
Lodders 1996) that these species are depleted (``rained out'')
in the atmospheres of all but the youngest (hence, hottest) substellar objects
and are sequestered in condensed form below the photosphere (Burrows, Marley, and Sharp 2000;
Marley \etal\ 1996; Allard \etal\ 1996; Tsuji \etal\ 1996).

However, the direct effect of clouds on the emergent fluxes of 
EGPs and brown dwarfs in the important T$_{\rm eff}$ range below 2000 K
has yet to be properly addressed.
Clouds can partially fill in spectral troughs and smooth spectral features (Jones and Tsuji 1997).
Furthermore, the presence or absence of clouds can strongly affect the reflection spectra and albedos of EGPs.
In particular, when there are clouds at or above the photosphere, the albedo in the optical
is high.  Conversely, when clouds are absent, the albedo in the mostly absorbing atmosphere is low.
Condensed species in brown dwarf and EGP atmospheres range from ammonia ice in low temperature objects
to silicate and iron grains at high temperatures.  Clearly, the physical properties of the variety of SMO clouds  
({\it i.e.}, their particle modal size and distribution, the amount of condensate, and
their morphology and spatial extent) are important and central challenges for the brown dwarf theorist.  

Unfortunately, cloud particle sizes are not easily modeled and are a strong function of
the unknown  meteorology in SMO atmospheres.  Inferred particle sizes in
solar system giant planet atmospheres can guide brown dwarf and EGP models, though they range
widely from fractions of a micron to tens of microns.
Importantly, atmospheric dynamics can lead to dramatic changes in the
mean particle sizes. In particular, convective processes lead to
growth in the mean particle size.
In the case of water and magnesium silicates, the latent heat of
condensation increases the mean upwelling velocity and can
exaggerate these effects, as quantified by Lunine \etal\ (1989).
The simple model of the transport processes in magnesium silicate
clouds presented in Lunine \etal\ suggests particle sizes in the range of
10-100 \mic are possible by coalescence, much larger than the
micron--sized particles one would assume from simple condensation.

For a one-dimensional cloud model, the particle size, its shape
distribution, and the amount of condensate (for a given set of 
potentially condensable gas-phase constituents) are
sufficient to compute the effect of clouds at a given 
level on the atmospheric thermal balance,
if one knows the optical properties of the condensate itself. In the model of
Lunine \etal (1989), growth rates for particles and droplets were calculated at each
T/P level using analytic expressions for growth rates
(Rossow 1978).  The Lunine \etal model assumes that the atmospheric
thermal balance at each level is dominated by a modal particle size that is the
maximum attainable when growth rates are exceeded by the sedimentation, or
rainout rate, of the particles. Because the sedimentation 
rate increases monotonically with particle size,
this must happen at and above some particle size.  The
amount of condensate was assumed proportional to the vapor pressure at each
level. Such a criterion represents a rough estimate for particle size and
condensable in an environment in which upwelling (due, for example, to convection)
is not present. In convective environments, Lunine \etal 
altered the criterion for maximum particle size to be that at which the upwelling
velocity matched the particle sedimentation velocity. The condensable
material available at each level is no longer proportional to the vapor pressure
itself, because particles are wafted across levels by convection. Lunine \etal 
assumed the condensate load at each level had the altitude profile $e^{-z/H}$,
where $H$ is the scale height and $z$ the altitude. This is almost certainly an
overestimate (see below). The advantage of this model is that few parameters
need to be specified for a given brown dwarf or EGP: T,P, condensable species
and associated thermodynamic parameters, convective flux or mean upwelling
velocity (hence, mixing length), and ancillary physical constants.

Marley \etal (1999) modified the model described above to allow for a particle
size distribution, not calculated explicitly but assuming a Gaussian about some
modal particle size, and explicitly expressed the condensate density in terms of
the amount of vapor exceeding the saturation vapor pressure at a given altitude.
Because particles and droplets have surface tension, a real atmosphere on the
verge of forming clouds will always have some modest supersaturation. Vapor
pressures at the cloud base may range from 1.01 times the saturation vapor pressure
(for Earth) to as much as twice the vapor pressure under certain conditions in
cold planetary atmospheres. The amount of condensate at each level is then just
enough to bring the coexisting partial pressure of condensables to exactly the
vapor pressure. Ackerman and Marley (2000) have gone a step further by explicitly
including the mass balance between upwardly diffusing vapor and downward
sedimentation (rainout) of condensate. This allows an amount of
condensate at each level that is not tied  directly to an assumed value of the
supersaturation, which cannot be calculated a priori for a given atmosphere
(though it can be estimated). However, in the Ackerman and Marley (2000) model,
there is an adjustable parameter which is the mass-weighted
sedimentation velocity expressed relative to the mean convective velocity. 
While the model can be fitted to Jovian (for example)
cloud properties, the sedimentation parameter cannot be
specified a priori via the properties of the atmosphere. 

An extension of the above models in convective regions is to include plume
models to calculate upwelling velocities and condensate loading at
each level. A first effort at this with a simple one-dimensional moist entraining
plume model was made for brown dwarfs by Lunine \etal (1989), based on a
model for Jupiter's clouds by Stoker (1986). To extend any of the above
models to more than a single dimension will in fact require numerical plume
models, as has been done by Yair \etal (1995 ) for Jupiter. Once launched in that
direction, a number of parameters required for numerical convective cloud
models will need to be specified from auxiliary theoretical models or just
assumed, because they cannot be inferred directly from current observations
of L dwarfs, T dwarfs, or even Jupiter. Nonetheless, it is evident from the study of Jupiter
that heterogeneous cloud distributions, in the areal sense, may be the standard
state for EGP's under at least some conditions. In such circumstances, one-dimensional
cloud models may overestimate the effect of clouds in one sense
(by not accounting for emergent flux unmolested by clouds) and
underestimating the effect in others (by not allowing for a population of multi-scale-height
convective cloud masses).  In this regard, Tinney and Tolley (1999) and Bailer-Jones and Mundt (2000)
have recently seen photometric variations in red dwarfs and L dwarfs that are best
interpreted in the context of temporal variations in large patchy cloud structures.  Bailer-Jones and Mundt (2000)
claim that the later spectral types (Ls) are more likely to manifest such variations,
lending credence to the hypothesis that condensates (expected at lower temperatures),
not magnetic star spots (expected for earlier spectral types), are responsible. 

Despite the still primitive nature of cloud modeling,  
it is clear that the stacking order and importance of silicate/iron clouds between 1500 K and 2400 K,
chloride/sulfide hazes below $\sim$1000 K,
water clouds near 300 K, and ammonia clouds below 200 K, as depicted in Fig. \ref{profile:cloud},
are robust theoretical expectations.  Indeed, silicate/iron clouds characterize the L dwarfs.
Hence, it is becoming clear that the presence and consequences of clouds
are defining whole new classes of astronomical 
objects, making cloud physics a central and permanent
feature of SMO science and an important focus 
for future theoretical research.   

\section{Opacities}
\label{opac}

The ingredients needed to generate non--gray spectral, color, and evolutionary models
of brown dwarfs, EGPs, and very-low-mass stars are clear.  They include (a) equations of state for
metallic hydrogen/helium mixtures and molecular atmospheres ({\it e.g.}, Saumon,
Chabrier, and Van Horn 1995; Marley and Hubbard 1988), (b) scattering and absorption opacities for
the dominant chemical species, (c) an atmosphere/transfer code to calculate emergent spectra, temperature/pressure
profiles, and the positions of radiative and convective zones, (d) an algorithm for converting a grid
of atmospheres into boundary conditions for evolutionary calculations, (e) chemical equilibrium
codes and thermodynamic data to determine the molecular fractions,
(f) a ``stellar'' evolution code, and (g) a model for clouds and grain/droplet scattering and absorption.
Of these, one of the most problematic is the opacity database.
In this section, we summarize the most important gas-phase molecular and atomic opacities
employed for the theoretical calculation of SMO atmospheres and spectra.
We have not tried to be complete, but hope to provide newcomers to the field
with a little guidance as they establish their own theory toolboxes.

\subsection{Molecular Opacities}
\label{moleopac}

A series of databases of molecular line lists and strengths have recently become available that are derived from
theoretical calculations that employ a variety of quantum mechanical methods
(Polyansky, Jensen, and Tennyson 1994; Wattson and Rothman 1992; Partridge and Schwenke 1997).
Using partition functions, local-thermodynamic-equilibrium (LTE) level densities, stimulated emission
corrections, and broadening algorithms, one generates opacity tables for
the spectral and atmosphere calculations.
For gaseous H$_2$O, Partridge and Schwenke have calculated the strengths of more than $3\times 10^8$ lines.
This database is still undergoing revision and other groups are about to release their own water
lists (U. J\o{rgensen}, private communication).
For other molecular species ({\it e.g.}, CH$_4$, NH$_3$, H$_2$O, PH$_3$, H$_2$S and CO),
the HITRAN (Rothman \etal\ 1992,1997) and GEISA (Husson \etal\ 1994)
databases can be augmented with additional lines from theoretical
calculations and measurements (Tyuterev \etal\ 1994; Goorvitch 1994; Tipping 1990; Wattson and Rothman 1992; Strong \etal 1993;
Karkoschka 1994; L.R. Brown, private communication).  This results in databases
for CH$_4$ of $1.9\times10^6$ lines, for CO of $99,000$ lines, for NH$_3$ of $11,400$
lines, for PH$_3$ of $11,240$ lines, and for H$_2$S of $179,000$ lines.
At higher temperatures than usually encountered in SMO atmospheres,
continuum opacity sources might include those due to $\rm H^-$ and $\rm H_2^-$.
Collision--induced absorption (CIA) by H$_2$ and helium (Borysow and
Frommhold 1990; Zheng and Borysow 1995; Borysow, J\o{rgensen}, and Zheng 1997) is 
a major process in the dense atmospheres of old brown dwarfs, EGPs, and Jovian planets
and is an increasing function of pressure.

Rudimentary datasets from which to derive FeH and CrH opacities 
can be derived using Schiavon \etal (1997), Schiavon (1998),
Phillips and Davis (1993), and Ram, Jarman, and Bernath (1993).  Rages \etal\ (1991) provide a convenient
formalism for Rayleigh scattering, important at shorter wavelengths.

Figure \ref{opac1500} depicts the absorption cross section spectra per molecule
of H$_2$O, CH$_4$, H$_2$, CO, and NH$_3$ at 2000 K and 10 bar pressure
from 0.5 \mic to 5 \mic.  Since it portrays the basic spectral features
of some of the dominant gas-phase species in EGP and brown dwarf
atmospheres, this is a useful figure to study.
The importance of water in defining the $Z$, $J$, $H$, $K$,
and $M$ bands is manifest, the major features of methane stand out (in particular the
bands at 1.7 \mic and 2.1 \mic and the
fundamental band at 3.3 \mic), the 2.3 \mic and 4.7 \mic features of CO
are clear, the broad H$_2$ peak at 2.2 \mic is prominent, and the major
NH$_3$ bands in the near-IR are shown (Saumon \etal 2000).
NH$_3$ has its strongest band near 10.5 \mic and, despite the fact that nitrogen (NH$_3$)
is not as abundant as oxygen (H$_2$O), this is where one should look
for the most distinctive ammonia signature.

Due to its importance in SMO spectral modeling, particluarly at 1.7 $\mu$m and 2.1$\mu$m,
the paucity of data for CH$_4$ (which has $>$one billion lines) is a major remaining
concern.  In particular, the absence in the existing CH$_4$ databases
of its hot bands can result in a dependence of the CH$_4$ opacity upon temperature with the wrong {\it sign}.
Given the available databases (lacking as they do these hot bands),
the resulting errors in the differential effect of methane at $H$ and $K$ can
severely compromise the interpretation of the infrared colors and effective temperatures of T dwarfs.
Clearly, this deficiency in the CH$_4$ database is an obstacle along the
road to a more precise theory of brown dwarf spectra and colors.
However, the fundamental band of methane at 3.3$\mu$m is reasonably well quantified.

In the case of TiO, one good source of line data is Plez (1998).
Another is J\o{rgensen} (1997),
where data are given for the 7 following electronic systems:
$\alpha$(C$^3\mydelta - $X$^3\mydelta$), $\beta$(c$^1\Phi - $a$^1\mydelta$),
$\gamma'$(B$^3\Pi - $X$^3\mydelta$), $\gamma$(A$^3\Phi - $X$^3\mydelta$),
$\epsilon$(E$^3\Pi - $X$^3\mydelta$), $\delta$(b$^1\Pi - $a$^1\mydelta$), and
$\phi$(b$^1\Pi - $d$^1\Sigma$).  This dataset contains in
total over 2.5 million lines, or nearly 13 million lines when isotopically
subsituted molecules are counted.
Each system file lists the line wavenumbers
in cm$^{-1}$ (note that 10$^4$ cm$^{-1}$ $\equiv$ 10$^4$\AA), 
in order of increasing wavenumber, the excitation energies
above the ground state in cm$^{-1}$, the $gf$ values of the spectral
lines, various indices identifying the quantum states of the
participating levels, and the shift in cm$^{-1}$ for the lines due to the
isotopically substituted molecules $^{46}$Ti$^{16}$O, $^{47}$Ti$^{16}$O,
$^{49}$Ti$^{16}$O and $^{50}$Ti$^{16}$O relative to the most abundant
isotopic form $^{48}$Ti$^{16}$O.  Seventy-four percent of titanium
is in the form of $^{48}$Ti, with the other four isotopes making up the
rest with fractions ranging by less than a factor of two, so the isotopic
versions have to be considered.
For VO, a line list provided by Plez (1999) is quite up to date and includes
the A-X, B-X, and C-X systems.  Because
$^{51}$V is by far vanadium's most abundant isotope, the lines of isotopically
substituted molecules are not necessary.
This list consisted of over 3 million lines ordered in decreasing wavenumber from
25939 cm$^{-1}$ to 3767 cm$^{-1}$, with $gf$ values being given together with
the excitation energy of the lower state, the vibrational and rotational
quantum numbers of the participating states, and the identity of the branch
(P, Q, or R).  (Figure \ref{opac.sum} in \S\ref{spectra} shows a representative opacity spectrum
dominated in the optical and near-IR by TiO and VO (mostly TiO).)

\begin{quote}
\begin{flushleft}
{\it When you have eliminated the impossible, what ever\\
     remains, however improbable, must be the truth.\phantom{1234}}\\
         Sir Arthur Conan Doyle - The Sign of Four
\end{flushleft}
\end{quote}

\subsection{The Alkali Metal Lines}
\label{shape}

The line lists and strengths for the neutral alkali elements (Li, Na, K, Rb and Cs) can be obtained
from the Vienna Atomic Line Data Base (Piskunov \etal 1995).
The general line shape theories of Dimitrijevi\'{c} and Peach (1990),
Nefedov \etal (1999), and Burrows, Marley, and Sharp (2000, hereafter BMS) 
can be used to calculate the alkali metal line core and far-wing opacity profiles.
However, more modern calculations of these profiles are sorely needed.
As stated in \S\ref{determ}, the optical and very near-IR spectra of L and T dwarfs
might be determined in large measure by the shapes of the wings of the K I resonance doublet and the Na D lines.
There is not another branch of astronomy
in which the neutral alkali line strengths 1000-3000 \AA\
from their line centers have ever before been of central concern.

For our discussion of the alkali metal opacities, we follow closely the text
of BMS.  The major transitions of immediate
relevance are those that correspond to the Na D lines at 5890 \AA\ and the K I resonance
lines at 7700 \AA.  Given the high H$_2$ densities in brown dwarf
atmospheres, the natural widths (for Na D, $\sim$0.12 m\AA) and
Doppler widths of these lines are completely overwhelmed by collisional broadening.  However,
in general the line shapes are determined by the radial dependence of the difference of the perturber/atom potentials for
the lower and upper atomic states (Griem 1964; Breene 1957,1981)
and these are rarely known.  The line cores are determined by distant
encounters and are frequently handled by assuming a van der Waals interaction
potential with an adiabatic impact theory (Weisskopf 1933; Ch'en and Takeo 1957;
Dimitrijevi\'{c} and Peach 1990) and the line wings are determined by
close encounters and are frequently handled with a statistical theory (Holtzmark 1925; Holstein 1950).
The transition between the two regimes is near the frequency shift ($\Delta\nu$), or detuning, associated with
the perturbation at the so-called Weisskopf radius ($\rho_{\rm w}$), from which the collision cross section employed in
the impact theory is derived (Spitzer 1940; Anderson 1950).  In the simple impact theory, the line core
is Lorentzian, with a half width determined by the effective collision frequency, itself
the product of the perturber density, the average relative velocity of the atom and the perturber ($v$), and
the collision cross section ($\pi\rho_{\rm w}^2$).  If the frequency shift ($\Delta\nu$) due to a single perturber
is given by $C_n/r^n$, where $r$ is the interparticle distance, then $\rho_{\rm w}$ is determined from the condition
that the adiabatic phase shift, $\int^{\infty}_{-\infty}2\pi\Delta\nu  dt$, along a classical straight--line trajectory, with
an impact parameter $\rho_{\rm w}$, is of order unity.  This yields $\rho_{\rm w}\propto (C_n/v)^{1/(n-1)}$.
For a van der Waals force, $n=6$.  In the statistical theory, the line shape is a power law that
goes like $1/\Delta\nu^{\frac{n+3}{n}}$, and this is truncated (cut off) by an exponential
Boltzmann factor, $e^{-V_0(r_s)/kT}$, where $V_0(r_s)$ is the ground-state perturbation
at the given detuning.  The detuning at the transition between the impact and
statistical regimes is proportional to $(v^6/C_n)^{0.2}$ (Holstein 1950).

All this would be academic, were it not that for the Na/H$_2$ pair the simple theory in the core
and on the red wing is a good approximation (Nefedov, Sinel'shchikov, and Usachev 1999).
For the Na D lines perturbed by H$_2$, BMS obtain from Nefedov,
Sinel'shchikov, and Usachev (1999) a $C_6$ of $2.05\times10^{-32}$ $cgs$ and a transition
detuning, in inverse centimeters, of 30 cm$^{-1}(T/500 K)^{0.6}$, where $T$ is the
temperature.  For the K I resonance lines, BMS scale from the Na D line data, using
a $C_6$ of $1.16\times10^{-31}$ {\it cgs}, itself obtained from the theory of Uns\"old (1955).
This procedure yields a transition detuning for the 7700 \AA\ doublet of 20 cm$^{-1}(T/500 K)^{0.6}$.
Nefedov, Sinel'shchikov, and Usachev (1999) show that for a variety of perturbing
gases the exponential cutoff for the Na D lines can be (for temperatures of 1000--2000 K)
a few$\times 10^{3}$ cm$^{-1}$.  The difference between 5890 \AA\ and
7700 \AA, in inverse wavenumbers, is $\sim$4000 cm$^{-1}$
and that between 7700 \AA\ and 1.0 \mic is only 3000 cm$^{-1}$.  Hence, it is reasonable to
expect that the detunings at which the line profiles are cut off can be much larger
than the Lorentzian widths or the impact/statistical transition detunings of tens of cm$^{-1}$.
Since there is as yet no good formula for the exponential cutoff term, BMS assume that it is
of the form $e^{-qh\Delta\nu/kT}$, where $q$ is an unknown parameter.
Comparing with the examples in Nefedov, Sinel'shchikov, and Usachev (1999), $q$ may be
of order 0.3 to 1.0 for the Na/H$_2$ pair.  Without further information or guidance,
BMS assumed that it is similar for the K/H$_2$ pair, but stress that this algorithm is
merely an ansatz and that the development of a more comprehensive theory based on the true perturber potentials
is vastly preferred.

Nevertheless, in the context of this approach to the alkali line profiles, we
can derive opacities as a function of wavelength.  Figure \ref{opac.nak.wei} depicts the
abundance-weighted opacities of the neutral alkali metal lines at 1500 K and 1 bar.
As we will see in \S\ref{determ}, the wings of the K I resonance doublet at 7700\AA\ quite naturally
have the proper slope and strength to explain the T dwarf spectra shortward of 1.0 \mic.

\subsection{Grain Scattering and Absorption Opacities}

Scattering and absorption by grains is traditionally handled using Mie theory, or approximations to Mie theory.
Given a particle radius ($a_p$) and a complex index of refraction, Mie theory provides scattering
and extinction cross sections, as well as angular scattering indicatrices (though usually only an asymmetry
factor, $g = <\cos\theta>$, is needed).  Large particles are very
forward-scattering and this can significantly alter the
reflectivity of EGP clouds at altitude (Seager, Whitney, and Sasselov 2000).
In addition, large particles require an increasing
number of terms in an infinite series to describe these parameters
accurately.  While the cross sections and scattering asymmetry factors of
small-- to moderately--sized particles ($2\pi a_p/\lambda$ \sles 75) vary substantially with wavelength, these
variations are greatly reduced for larger spheres.  For these larger
particles, an asymptotic form of the Mie equations outlined
by Irvine (1965) is useful. Interpolation between the full Mie
theory and these asymptotic limits yields the parameters for large
particles.  However, inherent assumptions in the asymptotic form of the Mie equations
render them inadequate for the computation of the scattering cross sections in the
weak-absorption limit ($n_{imag}$ \sles $10^{-3}$), in which case
the geometric optics approximation can be invoked (Bohren and Huffman 1983).

The principal condensates in SMO atmospheres for which one
has traditionally needed Mie theory include NH$_3$(c),
H$_2$O(c), MgSiO$_3$ (enstatite), Mg$_2$SiO$_4$ (forsterite), Al$_2$O$_3$ (corundum), Na$_2$S, solid alkali chlorides, and iron.
Some of the relevant complex indices of refraction as a function of wavelength can be obtained from
Martonchik \etal (1984), Warren (1984), Begemann \etal (1997), and Dorschner \etal (1995).
From the latter, one can also obtain the complex indices of refraction for the entire
pyroxene (Mg$_{x}$Fe$_{(1-x)}$SiO$_3$) and olivine (Mg$_{2(1-x)}$Fe$_{2x}$SiO$_4$)
series. However, generally the complex indices of refraction at
all the required wavelengths and for all the suspected condensates are difficult
or impossible to find.  Moreover, the grains are probably mixtures of various
species, may be layered, and are probably not spherical.  Added to this is the fact that the particle
size distributions can not yet be estimated with accuracy, though we know that in
convective layers the particle sizes must on average be large ($a_p\ \sgreat\  1-10$\mic) (\S\ref{clouds}).

Nevertheless, a simple estimate of the total optical depth due to grains can be obtained using the
arguments of Burrows and Sharp (1999) or Marley (2000) and hydrostatic equilibrium.  Following Marley (2000),
one finds that
\begin{equation}
\tau_{\lambda} \sim 75\epsilon Q_{\lambda} \phi \Bigl(\frac{P_c}{{\rm 1\ bar}}\Bigr)\Bigl(\frac{10^5
{\rm cm\ s}^{-2}}{g}\Bigr)\Bigl(\frac{1\ \mu{\rm m}}{a_p}\Bigr)\Bigl(\frac{1.0\ {\rm gm\ cm}^{-3}}{\rho}\Bigr),
\label{taumie}
\end{equation}
where $ Q_{\lambda}$ is the Mie total extinction factor
(which for large $2\pi a_p/\lambda$ asymptotically
approaches 2.0), $P_c$ is the pressure
at the base of the cloud, $g$ is the gravity, and $\rho$ is the grain material density. $\phi$
is the product of the number mixing ratio ($f$) and
the ratio of the condensate molecular mass to the
mean molecular weight of the atmosphere.   The $\epsilon$ factor
attempts to account for the degree of supersaturation and may be between $\sim$0.01 and 1.0 (\S\ref{clouds}).
When we insert reasonable values for all these quantities, eq. (\ref{taumie}) reveals
that for the most abundant condensates ({\it e.g.}, water [$f \sim 10^{-3}$], enstatite [$f \sim 10^{-4.5}$],
iron [$f \sim 10^{-4.5}$]), $\tau_{\lambda}$ in the optical and near-IR
can be from a few to $10^5$.  Clearly, grains can be very important in SMO atmospheres.
Crudely, $\tau_{\lambda}$ is inversely proportional to mean particle size
and gravity and is directly proportional to the pressure at the cloud base.
Small particles are efficient scatterers; the choice
between 0.1 \mic and 10 \mic makes a non-trivial difference.  Low-gravity atmospheres can
have thicker clouds ({\it cf.}, Saturn).  Since condensation is more closely tied
to temperature than pressure, low-entropy atmospheres (those with higher pressures at a given
temperature) have thicker clouds.   This is most germane for low-mass,
old brown dwarfs and EGPs, making their study more nuanced and akin to the study of the
planets and moons of our solar system.

For the time being, we must accept ambiguities in the compositions, grain sizes, optical properties,
and spatial distributions of cloud particles that can play central roles in SMO
theory.  The spectral data themsleves may provide the key, since precision
transit (\S\ref{irradiation}) and spectral measurements are tantamount to
remote sensing.  Attempts have been made to improve the situation vis \`a vis theory (Lunine \etal 1989; Ackermann and Marley 2000;
Tsuji \etal 1996; Allard \etal 1997; \S\ref{clouds}) and, given a complete cloud/grain model, theory can speak eloquently
concerning the consequences.  However, these cloud models are simplifications, mere guides to the actual meteorology.
Clearly, much interesting work remains to be done concerning the physics of clouds
in brown dwarf and giant planet atmospheres.

\subsection{On the Appropriateness of LTE Treatments}

Given the fact that the study of SMO and EGP atmospheres and spectra is still in its
pioneering phase and that there remain significant gaps in our knowledge of various
important opacities ({\it e.g.}, CH$_4$, alkali metal line wings, grains), it is
inappropriate to go beyond LTE algorithms to non-LTE approaches in order to credibly explain
the incoming data and the atmospheres of these objects. Furthermore, given
the high densities in these high-gravity atmospheres, the collision rates are high enough to keep the level populations
in LTE to high accuracy (Schweitzer, Hauschildt and Baron 2000).  Note that
the densities in the solar atmosphere are 10$^{3-4}$ times lower
than in brown dwarf and EGP atmospheres, yet even there non-LTE effects are rarely in
evidence or required to explain what is observed.  Though non-LTE effects needn't be addressed
for the vast majority of the transitions and abundances in substellar atmospheres, UV
irradiation of close-in EGPs may create ionospheres in their upper atmospheres.  This is
particularly relevant for the neutral alkali metals whose opacities play such
an important role at depth and for SMOs in isolation.  Given this, the next phase of
theoretical modeling of close-in EGP atmospheres should involve
the ionization/recombination equations for the alkali metals and electrons.
Note that at the high densities of these atmospheres, three-body
recombination rates seem to dominate (Sudarsky, Burrows, and Pinto 2000).

\section{Brown Dwarf, L Dwarf, and T Dwarf Colors and Spectra}
\label{spectra}

The studies of Burrows \etal\ (1989,1993,1997), Tsuji \etal (1996,1999), Marley \etal\ (1996), Baraffe \etal (1998),
Chabrier and Baraffe (1997), and Allard \etal (1996) have revealed major new aspects of
EGPs and brown dwarfs that bear listing and that uniquely characterize them.
Generally, those molecules that dominate in abundance also dominate the opacity.
Hence, above $\sim$1500 K the dominant opacity sources are H$_2$O, CO, and 
silicate grains, to which above $\sim$2000 K are added gaseous TiO, and VO (\S\ref{moleopac}).
The major opacity sources for \teffs below $\sim$1500 K are 
H$_2$O, CH$_4$, NH$_3$, H$_2$, and the alkali metals (shortward of 1.0 \mic, \S\ref{shape});
these atmospheres are otherwise depleted of heavy elements (\S\ref{chem} and \S\ref{clouds}).
For T$_{\rm eff}$s below $\sim$500 K, water clouds form at or above the photosphere 
(signaling a new, as yet unidentified, ``water cloud'' spectral class)
and for T$_{\rm eff}$s below 200 K, ammonia clouds form ({\it viz.,} Jupiter).  Collision--induced absorption (CIA)
of H$_2$ and rotational transitions of H$_2$O and CH$_4$ partially suppress emissions longward of $\sim$10 \mic.  The holes in the opacity
spectrum of H$_2$O that define the classic telluric IR bands also regulate much of the emission from
EGP/brown dwarfs in the near infrared.  Importantly, the windows in H$_2$O and the suppression by H$_2$ conspire to
force flux to the blue for a given T$_{\rm eff}$ (Marley \etal 1996).
The upshot is an exotic spectrum enhanced relative to the black body value
in the $Z$ ($\sim$1.0 \mic), $J$ ($\sim$1.2 \mic), and $H$ ($\sim$1.6 \mic) bands
by as much as {\it two} to {\it five} orders of magnitude and $J-K$ and $H-K$ colors
that become {\it bluer}, not redder, with decreasing \teff (Burrows \etal 1997).
Figure \ref{gl229b.fit.5} portrays the near-infrared spectrum of 
Gl 229B (along with a representative theoretical fit)
and demonstrates these characteristics quite well.  
Therefore, the dominance of water in both SMO and terrestrial atmospheres is fortuitous
for ground-based observations; the emission peaks in SMOs naturally coincide in wavelength
with the telluric atmospheric windows.

Figure \ref{spec.1mj} portrays the low-resolution evolution of the spectrum of a cloudless 1 \mj object and compares
it to the corresponding black body curves.  This figure (along with Fig. \ref{gl229b.fit.5}) 
demonstrates how unlike a black body an SMO spectrum
is.  The enhancement at 5 \mic\ for a 1 Gyr old, 1 \mj extrasolar planet is by four orders of magnitude.
As T$_{\rm eff}$ decreases below $\sim$1000 K, the flux in the $M$ band ($\sim$5 \mic)
is progressively enhanced relative to the black body value.
While at 1000 K ( {\it cf.}, Gl 229B) there is no enhancement, 
at 200 K it is near 10$^5$ for cloudless atmospheres.  Therefore, the $Z$, $J$, $H$, and $M$ bands
are the premier bands in which to search for cold substellar objects.
Even though $K$ band ($\sim$2.2 \mic) fluxes are generally higher
than black body values, H$_2$ and CH$_4$ absorption features in the $K$ band decrease its importance
{\it relative} to $J$ and $H$.  In part as a consequence of the increase of atmospheric
pressure with decreasing T$_{\rm eff}$, the anomalously blue $J-K$ and $H-K$
colors get {\it bluer}, not redder.  

However, the shape of the H$_2$O absorption 
spectrum plays the major role in the infrared blueness of brown dwarfs shortward of 2.5 \mic (\S\ref{moleopac}).
Figure \ref{opac.sum} demonstrates this clearly.  Depicted are composition-weighted opacity spectra at 2200 K (including
TiO and VO) and at 1000 K (after TiO, VO, and silicates have rained out).  The former is representative
of M dwarfs and shows why they are red below $\sim$2.0 \mic (due to TiO and VO) and the latter is representative of T dwarfs
and shows why they are blue in the same spectral range (due to H$_2$O).  Methane absorption at $\sim$1.7 \mic 
and $\sim$2.15 \mic contributes further to this trend by shaving the red sides of the $H$ and $K$ bands
in ways that have come to characterize and define the T dwarfs.  
Note that Noll \etal (2000) have recently identified the 3.3 \mic methane feature (the $\nu_3$ band) in some 
L dwarfs, so the presence of methane spectral features alone does not 
distinguish a T dwarf.  A T dwarf is defined by the appearance of methane in the
$H$ and $K$ bands, not the $L$ band (Burgasser \etal 2000).
Nevertheless, methane is not a factor in stars, but it is
a distinguishing feature of giant planets such as Jupiter and Saturn.
Between the M and T dwarfs, the L dwarfs represent the transition subtype
in which silicate and iron clouds form in their atmospheres, wax in importance, and then wane to turn
into heavy-element-depleted T dwarfs with blue infrared colors.  

\begin{quote}
\begin{flushleft}  
{\it It is a capital mistake to theorize\\
before one has data.\phantom{12345678901}}\\
         Sir Arthur Conan Doyle - Scandal in Bohemia  
\end{flushleft}  
\end{quote}

\subsection{Observed L and T Dwarf Properties}
\label{observed}

The first L dwarf was actually discovered by Becklin and Zuckerman (1988)
as a resolved companion (GD 165B) to the white dwarf GD 165 (see \S\ref{edge}).
Kirkpatrick, Henry, and Liebert (1993) found that 
its spectrum differed decisively at red wavelengths,
which ultimately led to the 
establishment of the L spectral class. 
Additional L dwarf companions are listed in Table \ref{companions} 
(\S\ref{pop} and Reid \etal 2000b).  However, their
counterparts in isolation are much more numerous. 
To date, based on selection by red colors, 2MASS (Kirkpatrick
\etal 1999,2000), the Sloan Digital Sky Survey (Fan \etal 2000),
and DENIS (Delfosse \etal 1997; Martin \etal 1999) have found more than 150 L dwarfs in the field.

Figure \ref{liebert.A} shows red (0.6-1 \mic) spectra of a late M dwarf and three L dwarfs that
span the L spectral sequence.
L dwarfs may have \teffs between 1300 K and 2100 K (\S\ref{empirical}) and, as indicated in Fig. \ref{liebert.A},
are characterized by the clear onset of metal oxide (TiO and VO) depletion (Jones and Tsuji 1997),
the formation of iron and silicate refractories,
the appearance of metal hydrides (FeH and CrH), and the dramatic growth in strength of
the neutral alkali metal lines of potassium, lithium, cesium, rubidium, and sodium
in their optical and near-infrared spectra (Martin \etal 1999; Kirkpatrick \etal 1999,2000).
Lines of Cs I at 8521 \AA\ and 8943 \AA,
Rb I at 7800 \AA\ and 7948 \AA, Li I at 6708 \AA, and Na I at 
5890/5896 \AA\ (Na D), 8183/8195 \AA, 1.14 \mic, and 2.2 \mic have been
identified in L dwarf spectra.  Importantly, as Fig. \ref{liebert.A} clearly shows, the K I doublet at 7700 \AA\ 
emerges to dominate in the spectra of mid- to late L dwarfs where in earlier types TiO/VO once held sway.   
L dwarfs such as DENIS-1228, 2MASS-0850,
2MASS-1632 (Kirkpatrick \etal 1999), and Denis-0205 (Leggett \etal 2000b)
are particularly good examples of this fact.
Note that both DENIS-1228 and Denis-0205 are doubles (Table \ref{companions}; Leggett \etal 2000b; Reid \etal 2000b).

Figure \ref{1507} shows the Keck II spectrum of the 
bright L5 dwarf,  2MASSW J1507, from 4000\AA\ to 10000\AA\ (Reid \etal 2000a).  The Na I
resonance doublet is even more dominant than the K I doublet over these
wavelengths, presumably due to sodium's higher abundance.
The Reid \etal (2000a) spectrum of 2MASSW J1507 also suggests
that the strong lines of K, Cs, and/or Rb near $\sim$0.4 \mic
might soon be identified.  Those lines are seen in Fig. \ref{opac.nak.ind},
which portrays the full cross-section spectrum per molecule shortward of 1.5 \mic and at
T = 1500 K and 1 bar of the neutral alkali metals.  Included in this figure are all the subordinate lines
excited at this temperature.

The L dwarfs constitute the spectroscopic
link between M dwarfs and Gl 229B-like objects (T dwarfs).  
T dwarfs differ from L dwarfs primarily in their infrared spectra and
colors and all T dwarfs are brown dwarfs.  They have stronger H$_2$O and H$_2$ absorption, while strong
CH$_4$ bands (in $H$ and $K$) appear in place of the CO bands (particularly at 
2.3 \mic and 4.7 \mic) seen in M dwarfs.  Figure \ref{liebert.C}
shows the infrared spectra of a late L and two T dwarfs, the later being
the prototypical Gl 229B.  Superposed are the nominal response curves
for the $J$ (1.2 $\mu$m), $H$ (1.6 $\mu$m) and $K$ (2.2 $\mu$m) photometric bands.  It
is apparent, for the reasons mentioned above, that
the $J-K$ color of a T dwarf is blue, like an A star.  
This point is made even more forcefully in Fig. \ref{JJK},
a M$_J$ versus $J-K$ color-magnitude diagram.  Included are various
M dwarfs, L dwarfs, Gl 229B, theoretical isochrones, and the corresponding
black body isochrones.  The extremely blue deviations  
of Gl 229B, in particular, and T dwarfs, in general,
from black-body values are starkly clear.

SDSS 1021 on Fig. \ref{liebert.C} is a
``transition" L/T object, exhibiting both CH$_4$ and CO absorption in the
$H$ and $K$ passbands (Leggett \etal 2000a).  Until very recently,
a gap in the infrared colors ({\it e.g.}, $J-K$) of the discovered L dwarfs and T dwarfs existed.
The $J-K$ color of Gliese 229B is blue ($\sim -0.1$) , while that of the latest L dwarfs
is red ($\sim 2.1$; Kirkpatrick \etal 1999).  However, within the last year,
putative ``missing links'' with $J-K$ colors of $0.8$--$1.5$ have been discovered in this gap
(Leggett \etal 2000a; Table \ref{adam}), signaling the depletion and rainout
of refractories below the photosphere (Ackerman and Marley 2000),
the appearance of methane absorption bands near the $H$ and $K$ bands, and the
increasing importance of collision-induced absorption (CIA) by H$_2$ (\S\ref{moleopac}).  The latter is
crucial to the understanding of the atmospheres of Jupiter and Saturn, a fact which serves to emphasize that these new
populations bridge the planetary and stellar domains.
Figure \ref{liebert.D} demonstrates the dramatic jump in the $JHK$ two-color plot made
between the latest L and T dwarfs, with L/T transition objects falling
in between.  M dwarfs are plotted as x's, L dwarfs as filled circles, L/T
and early T dwarfs as open squares, and Gl 229B-like T dwarfs as open
triangles.  

Young cluster brown dwarfs can have effective temperatures and luminosities in the stellar range and be spectroscopically M type
(Martin \etal 1996; Zapatero-Osorio, Rebolo, and Martin 1997; Luhman \etal 1998; Luhman 1999 )
and old brown dwarfs can cool to achieve the temperatures one might associate with a T dwarf, EGP, or young Jupiter.
Figure \ref{teff} depicts the evolution of \teff versus age for solar-metallicity SMOs and stars from 0.3 \mj
to 0.2 \mo ({\it cf.} Fig. \ref{lum.ps}) and demonstrates this dichotomy quite well.
A young brown dwarf can begin with M dwarf temperatures,
then age into the L dwarf regime, and end up after significant evolution as a T dwarf.
Hence, as Fig. \ref{teff} shows, whether a given object
is an M, L, or T dwarf depends upon its mass {\it and age}.

\begin{quote}
\begin{flushleft}  
{\it The wind and the waves are always on the side\\
of the ablest navigators.\makespace\phantom{123}}\\
         Gibbon - Decline and Fall of the Roman Empire
\end{flushleft}  
\end{quote}

\subsection{Empirical Temperature Scales}
\label{empirical}

The substantial number of field L dwarfs with trigonometric parallax
determinations -- along with several L and T objects with stellar
companions of known distance -- yield the H-R diagram shown in Fig. \ref{liebert.E}.
The two T dwarfs (Gl 229B and Gl 570B) are wide companions to stars with
well-determined distances.  The prototype, Gl 229B, has M$_J$=
15.4$\pm$0.1.  Gl 570D is a companion to a nearby triple star and 
at M$_J$= 16.47$\pm$0.05 is the
faintest known T dwarf (Burgasser \etal 2000).
Interestingly, the faintest L dwarfs with parallaxes on Fig. \ref{liebert.E}
have M$_J$s only half a magnitude brighter than Gl 229B (or one magnitude
brighter at M$_K$, due to the color change).  Since detailed, multi-wavelength
analyses ({\it e.g.}, Marley \etal 1996) suggest that Gl 229B has a \teff near 950 K, the
difference in luminosity between Gl 229B and Gl 570D indicates that
Gl 570D has a \teff near 800 K (assuming a similar radius).

Only rough estimates of \teff currently exist for L dwarfs, even though 
observed changes in spectral features led to detailed spectral
classification systems for them (Kirkpatrick \etal 1999; Martin \etal 1999).
Since the L dwarf spectral types increase monotonically
with color and luminosity (Fig. \ref{liebert.E}), there is every reason to believe
that they represent essentially a rank ordering with decreasing \teff,
as is the case for M dwarfs.  However, at issue are the range, the maximum \teff, 
and the minimum \teff of L dwarfs.

Kirkpatrick \etal (1999, 2000) have used the appearance/disappearance
of individual spectral features such as TiO, VO and CH$_4$ as the basis
for a scale running from 2000/2100 K (type L0) to 1300/1400 K for L8
dwarfs.  (For the remainder of this discussion, we adopt the spectral types
of Kirkpatrick \etal .)  The top of the range is set by the
weakening of TiO and VO, as discussed in \S\ref{chem}, while the
appearance of CH$_4$ predicted near the bottom of this range should
signal the end of L dwarfs and the L/T transition.  In contrast,
Basri \etal (2000) prefer 
temperatures determined from fitting high-resolution alkali line profiles
with model atmospheres.  They derive a hotter scale of 2200 K to
1700 K.  In favoring a 700 K ``gap" between the coolest known L dwarf and
Gl 229B, they argue that later L subtypes and many L/T transition
objects populate the gap.  However, Pavlenko, Zapatero-Osorio, 
and Rebolo (2000) fit the 0.65 to 0.9 \mic spectrum (including 
the alkali metal lines) of the very late L dwarf DENIS-0205  
to models with \teffs between 1200 K and 1400 K,
depending on the broadening theory employed, thus favoring the low-temperature scale.

There are good reasons to argue that the actual gap between L8 and T
dwarfs is much smaller than 700 K.  Figure \ref{liebert.C} shows that, unlike the $H$
and $K$ bands, the $J$ band is relatively unaffected by the transition from
late L to T.  Therefore, one might expect the ordinate of Fig. \ref{liebert.C} to
provide a reliable measure of M$_{bol}$.  In astronomical language, this is
equivalent to making the assumption that the bolometric correction (BC$_J$) between
M$_J$ and M$_{bol}$ (the total luminosity) is varying slowly between
late L and T.  Multiwavelength observations of Gl 229B show that it has
a BC$_J$ of 2.2 magnitudes, and, hence, M$_{bol}$ = 17.7.  However, even
late M dwarfs have BC$_J$ values only 0.2 mags different, so the above
assumption is probably safe.  Hence, the close proximity of late L and T
dwarfs in M$_J$ values is evidence that the temperature gap is small.
It is likely that L/T transition objects populate this temperature
region.  For a more detailed development of these arguments, the reader is referred
to Reid (2000). 

In contrast to the L and M dwarfs, T dwarfs so far show relatively
modest variations in infrared spectra 
(Burgasser \etal 1999; Strauss \etal 1999; Tsvetanov \etal 1999), 
except for the L/T transition objects (Leggett \etal 2000a).  It is
thus a more daunting task to set up a spectral classification scheme
which can rank-order and help determine their effective temperatures and 
luminosities.

\subsection{Gliese 229B as a T Dwarf Benchmark and the Role of the K and Na Resonance Lines}
\label{determ}

As we have stated, the discovery of the T dwarf, Gliese 229B, in 
1995 and at 5.8 parsecs was a milestone in the study of brown dwarfs,
providing the first bona fide object with an effective temperature 
(T$_{\rm eff} \sim$ 950 K)  and luminosity ($\sim 7\times 10^{-6}$ \lo) that
were unambiguously substellar (Nakajima \etal 1995; Oppenheimer \etal 
1995,1998; Marley \etal 1996; Allard \etal 1996;
Geballe \etal 1996; Schultz \etal 1998; Saumon \etal 2000).
Since Gl 229B, {\it twenty-four} similar substellar-mass objects
have been discovered in the field by the Sloan Digital Sky Survey (SDSS, Strauss \etal 1999; Tsvetanov \etal 1999; Leggett \etal 2000a),
by the 2MASS survey (Burgasser \etal 1999, 2000), and by the NTT/VLT (Cuby \etal 1999) (see Table \ref{adam} for a subset).
These brown dwarfs have spectral
differences from 0.5 \mic to 5.0 \mic that reflect true differences
in  gravity, \teff, and composition (metallicity), from which, with the aid
of an evolutionary code, the radius, mass, age, and luminosity can be derived.
The spectrum of Gl 229B depicted in Fig. \ref{gl229b.fit.5} is typical of the new T dwarf class.
It is characterized by emission spikes through the brown dwarf's water absorption bands
at $Z$, $J$, $H$, and $K$, strong absorption at the 3.3 \mic feature of methane,
the characteristic methane feature at 1.7 \mic on the red side of the $H$ band, and 
alkali metal absorption lines shortward of one micron.  

The importance of the neutral alkali metals (\S\ref{chem} and \S\ref{shape}) led Burrows, Marley, and Sharp (2000, BMS) 
to conclude that the strong continuum absorption seen
in all T dwarf spectra in the near-infrared from 0.8 \mic to 1.0 \mic, previously interpreted as due to
an anomalous population of red grains (Griffith, Yelle, and Marley 1998) or in part due to
high-altitude silicate clouds (Tsuji \etal 1999; Allard \etal 1997), is
in fact most probably due to the strong red wings of the K I doublet at $\sim$7700 \AA.
Figure \ref{gl229b.K} from BMS demonstrates the naturalness 
with which the potassium resonance lines alone fit
the observed near-infrared/optical spectrum of Gl 229B (Leggett \etal 1999; though note the caveats in \S\ref{shape}).
In Fig. \ref{gl229b.K}, four theoretical models with about the same luminosity and core entropy (roughly satisfying
the relation $M/t^{1/2}\sim$ const.) are compared to the  Leggett \etal (1999) 
calibration of the Gl 229B data shortward of 1.45 \mic.
Curiously, in the metal-depleted atmospheres of T dwarfs the reach of the K I doublet
is one of the broadest in astrophysics, its far wings easily extending more 
than 1500 \AA\ to the red and blue.  With rainout, below
$\sim$1000 K sodium and potassium exist as sulfides or chlorides (Na$_2$S and KCl/K$_2$S) (Lodders 1999).   Without rainout,
complete chemical equilibrium at low temperatures requires that sodium and potassium reside in
the feldspars.  If such compounds formed and persisted at altitude, then the
nascent alkali metals would be less visible, particularly in T dwarfs.  By modeling spectra with and
without the rainout of the refractories and comparing to the emerging library of T dwarf spectra ({\it e.g.}, Burgasser \etal 1999,2000;
Leggett \etal 2000a; McLean \etal 2000), the degree of rainout and the composition 
profiles of brown dwarf atmospheres can be approximately ascertained.

The 1.17 \mic and 1.24 \mic subordinate lines of excited K I have also been identified in T dwarfs
(Strauss \etal 1999; Tsvetanov \etal 1999; McLean \etal 2000).  Since these subordinate lines are on the crown of
the $J$ band, they allow one to probe the deeper layers at higher temperatures.  Figure \ref{bright}
portrays for a representative Gl 229B model the dependence on wavelength of the ``brightness'' 
temperature, here defined as the temperature at which the photon optical depth ($\tau_{\lambda}$) is $2/3$.
Such plots clearly reveal the temperature layers probed with spectra and 
provide a means to qualitatively gauge composition profiles.  
Specifically, for the Gl 229B model the detection of the subordinate lines
of potassium indicates that we are probing to $\sim$1550 K, while the detection of the fundamental methane
band at $3.3$ \mic means that we are probing to only $\sim$600 K.     

As Fig. \ref{gl229b.K} suggests, the BMS theory also explains the WFPC2 $I$ band (M$_I\sim$20.76; theory = 21.0)
and $R$ band ($M_R \sim 24.0$; theory = 23.6) measurements made of Gl 229B (Golimowski \etal 1998),
with the Na D lines at 5890 \AA\ helping to determine the strength of the $R$ band.
BMS predicted not only that there would be a large trough in
a T dwarf spectrum at 7700 \AA\ due to the K I resonance,
but that the spectrum of a T dwarf
would peak between the Na D and K I absorption troughs at 5890 \AA\ and 7700 \AA, respectively. This prediction
was recently verified by Liebert \etal (2000) for the T dwarf SDSS 1624+00 and recapitulates the 
behavior seen in Fig. \ref{1507} for late L dwarfs.

Since in brown dwarfs the distributions and relative depths of the alkali metals 
depend systematically upon \teff and $g$,
the alkali metal lines, along with H$_2$O, CH$_4$, and H$_2$ bands in the near infrared,  
may soon be used to probe the atmospheric structure of T dwarfs.
Marley \etal (1996) and Allard \etal (1996) analyzed the full
spectrum of Gl 229B from the $J$ band ($\sim$1.2 \mic) through the $N$
band ($\sim$10 \mic), but could constrain its gravity to within no better than a factor
of two.  Such gravity error bars are large and translate into
a factor of $\sim$2 uncertainty in Gl 229B's inferred mass and age.
However, with the new Keck, UKIRT,
HET, and VLT observations (for example) and our emerging understanding of what determines the optical and infrared
spectra, we can now hope to obtain tighter constraints on T dwarf gravities, and, hence, masses.
In particular, due to the differential dependence on pressure of the CH$_4$ and H$_2$ opacities,
as well as the shifting competition between the H$_2$O/H$_2$/CH$_4$ band strengths and the
alkali line strengths with changing \teff and $g$, the near IR spectra from 1.1 \mic to 2.3 \mic, the $H-K$ color, and the continuum shape
from 0.7 \mic to 0.95 \mic can be used to break the degeneracy in the \teff/gravity/composition fits.
What is being revealed is that the T dwarfs discovered to date span a wide range in age, 
in mass (perhaps $\sim$20 \mj to $\sim$70 \mj), and
in \teff (perhaps $\sim$700 K to $\sim$1200 K). 

\subsection{The Color of a ``Brown'' Dwarf}
\label{color}

Curiously, since the Na D lines can be prominent in brown dwarf spectra and
will suppress the green wavelengths and since the color ``brown'' is two parts red, one
part green, and very little blue, brown dwarfs should not be brown.  In fact,
recent calculations suggest that they are red to
purple, depending upon the exact shape of the line wings of Na D, the abundance
of the alkalis, the presence of high-altitude clouds, and the role of water clouds
at lower \teffs ($\sles$ 500 K).  A mixture of red and the complementary color
to the yellow of the Na D line makes physical sense. It is the {\it complementary} color,
not the {\it color}, of the Na D line(s) because Na D is seen in absorption, not emission.
Indeed, the recent measurement of the spectrum of the
L5 dwarf 2MASSW J1507 from 0.4 \mic\ to 1.0 \mic (Fig. \ref{1507}; Reid \etal 2000a)
indicates that this L dwarf is magenta in (optical) color.
This is easily shown with a program that generates the RGB equivalent of a given optical spectrum
(in this instance, R:G:B::1.0:0.3:0.42, depending upon the video ``gamma'').
Hence, after a quarter century of speculation and ignorance,
we now have a handle on the true color of a brown dwarf --- and it is not brown.

\section{Population Statistics, the Substellar IMF, and Brown Dwarf Companions}
\label{pop}

As is becoming increasingly clear, brown dwarfs are prevalent in  
the galactic disk, though they contribute little to
the galactic mass budget.
Not only are isolated brown dwarfs numerous in the solar
neighborhood, they are also numerous in Galactic disk clusters.
Many brown dwarfs have been found down
to masses near or below 10 \mj in the young $\sigma$ Ori cluster (Zapatero-Osorio
\etal 2000).  In the Pleiades, brown dwarfs have been found down to as far as 
$\sim$0.035 \mo ($\sim$35 \mj) (Martin \etal 1998).  Bouvier \etal 
(1998) find that the Pleiades IMF is still rising in the
substellar domain, with a slope consistent with an IMF ($dN/dM$) that is proportional to $M^{-0.6}$.  Indeed,
for several very young clusters spanning nearly two orders of magnitude
in stellar density (IC348, L1495E, $\rho$ Oph, and the Orion Nebula
Cluster),  Luhman \etal (2000) determine substellar IMFs that are flat or
slowly rising with decreasing mass.  Furthermore, the initial assessment of the L dwarf
sample from 2MASS suggests that the field mass function could be similar to
those of the clusters (Reid \etal 1999).

Of particular importance is the discovery of brown dwarfs in binary
systems.  Evidence has been accumulating that the binary frequency
declines with decreasing stellar mass, from the 60\% or greater
for solar-type stars (Duquennoy and Mayor 1991) to $\sim$35\% for M
dwarfs (Fischer and Marcy 1992; Reid and Gizis 1997).  Substellar-mass
companions to solar-type stars have generally not been found in 
radial velocity surveys; this is the so-called ``brown dwarf desert'' (Marcy and
Butler 1998).  Several ground and space-based studies reach similar
conclusions: M dwarfs do not seem to harbor close brown dwarf
companions.  However, brown dwarfs are not so uncommon as wide
companions ($\sgreat$10 A.U.) over a broad range of spectral type (G--M) and mass (Kirkpatrick \etal 2001).
Likewise, the frequency of brown dwarf pairs is
appreciable -- perhaps 20\% (Reid \etal 2000b).  

The known brown dwarfs in binaries 
are listed in Table \ref{companions}.  They include the spectral prototypes GD 165B
and Gl 229B, and the low-luminosity T dwarf Gl 570D.  Also noteworthy 
is the first binary brown dwarf to be recognized -- PPL 15 in the
Pleiades cluster.  The others are a mix of brown dwarf companions to
known stars, binary brown dwarfs (found in the DENIS and 2MASS
surveys), and those found in proper motion studies.  Notice that the
mass ratios vary from essentially unity for brown dwarf pairs -- an
obvious selection effect -- to very small values (for companions to G
stars).  However, all pairs with separations less that 40 A.U. are
likely brown dwarfs with similar masses. These pairs include 
potential targets for astrometric follow-up. 

Population studies
of brown dwarfs in binaries, as well as in isolation, are 
yielding a rich harvest of information concerning not only the IMF
below the main sequence edge, but the process of ``star'' formation itself.  
The different binary fractions and orbital-distance distributions
above and below the HBMM speak directly to
the processes by which brown dwarfs themselves form.  Nevertheless, however they form,
brown dwarfs are significantly more abundant than most astronomers believed
even one year ago (circa 2000).  This is quite gratifying to those of us
who have been generating the associated theory in quiet anticipation.

\begin{quote}
\begin{flushleft}
{\it Then felt I like some watcher of the skies\\
     When a new planet swims into his ken;\\
     Or like stout Cortez when with eagle eyes\\
     He stared at the Pacific -- and all his men\\
     Look'd at each other with a wild surmise --\\
     Silent, upon a peak in Darien.}\\ 
         John Keats - On First Looking into Chapman's Homer
\end{flushleft}
\end{quote}

\section{New Worlds: Extrasolar Giant Planets}
\label{basics:egp}

Table \ref{beast} lists data on the EGPs discovered
as of August 2000 in order of increasing semi-major axis.  These discoveries have been made by 
a small, but growing, army of observers (Butler \etal 1997,1998; Cochran \etal 1997; Delfosse \etal 1998;
Fischer \etal 1999; Henry \etal 2000; Korzennik \etal 2000; Latham \etal 1989; 
Marcy and Butler 1996; Marcy \etal 1998; Marcy \etal 1999; Marcy, Butler, and Vogt 2000;
Marcy, Cochran, and Major 2000; Mayor and Queloz 1995; Mazeh \etal 2000; Noyes \etal 1997;
Queloz \etal 2000; Santos \etal 2000; Udry \etal 2000; Vogt \etal 2000; and references therein). Defined as objects  
found using the high-precision radial-velocity technique around stars 
with little or no intrinsic variability and previously thought to
be without companions, these 55 EGPs span more than two orders of magnitude
in $m_p \sin (i)$, semi-major axis, and period.  Collectively, EGPs show the wide range of eccentricities typical
of stellar companions, while those within $\sim$0.07 A.U. have the small eccentricities expected for objects
that have experienced significant tidal dissipation.   There is an interesting excess of primaries with
super-solar metallicities, as yet unexplained.  Multiple EGPs 
are known in a few systems ({\it e.g.}, 55 Cnc, $\upsilon$ And) and this number is sure to grow
as measurements extend to longer orbital periods and known velocity residuals are patiently followed. 

EGP surface temperatures (inferred from semi-major axes, 
stellar luminosities, ages, Bond albedos, and EGP masses) 
range from $\sim$200 K for distant EGPs to $\sim$1600 K for the close-in EGPs 
(Marley \etal 1999; Sudarsky, Burrows, and Pinto 2000; Table \ref{beast}; \S\ref{albedo}).   Though there is a selection
bias for close-in planets in short-period orbits and though ``Jupiters'' with longer periods (11.9 years)
would not as yet have been detected, the properties of this growing family of EGPs
could not be more unexpected.  However, we can not now distinguish
between true EGPs (planets) and brown dwarfs, as defined in \S\ref{over}.
The inclination angles for most of the radial-velocity objects are unknown; some ``EGPs''
are bound to be much more massive than Jupiter and to be brown dwarfs on the tail of a ``stellar'' 
population of companions.  Importantly, for semi-major axes less than $\sim$4.0 A.U., 
Table \ref{beast} indicates that there is a dearth of objects
with $m_p \sin (i)$s above 10 \mj (Marcy and Butler 1998); such a population would easily have been
detected if it existed (Marcy and Butler 1998).  This implies that most of the ``EGPs'' in Table \ref{beast} are in the
class of true planets distinct from higher-mass brown dwarfs and that the latter, if companions,
are preferentially found at larger orbital distances.

A histogram of $m_p \sin (i)$s does indeed imply that most of the 
EGPs in Table \ref{beast} are a population distinct from ``stars.''
Under the assumption that the inclination angles are randomly distributed on the
sky, one can derive the intrinsic mass distribution 
(ignoring selection biases!) from the observed distribution
using the formula:
\begin{equation}
\frac{d N}{d {m_p}} = \int^{m_p}_0 \frac{d N}{d {\rm m^{\prime}}} \frac{{\rm m^{\prime}}^2 d {\rm m^{\prime}}}
{ {m_p}^2 ({m_p}^2 - {\rm m^{\prime}}^2)^{1/2} } ,
\label{distrib}
\end{equation}
where $\frac{d N}{d {\rm m^{\prime}}}$ is the observed 
distribution of $m_p \sin (i)$s and $\frac{d N}{d {m_p}}$ is the 
intrinsic distribution of planet masses.  Equation (\ref{distrib}) shows that
$\frac{d N}{d {\rm m^{\prime}}}$ and  $\frac{d N}{d {m_p}}$ 
are not very different.  This can also be seen using   
the formula for the probability that an EGP mass 
is greater than $m$, given  $m_p \sin (i)$:
\begin{equation}
P(> m) = 1 - \sqrt{1-(m_p \sin (i)/m)^2} .
\label{prob}
\end{equation}
This formula indicates that there is only a 13.4\% chance 
for the planet mass to be greater than twice $m_p \sin (i)$ and that the average
possible planet mass, inferred from $m_p \sin (i)$, is only $\pi/2$ times bigger. 

There are many facets to the study of EGPs, both observational and theoretical,
in which an expanding circle of researchers are now engaged.  The subject is growing geometrically.
However, certain theoretical subtopics (such as transit, albedo, reflection spectra, and phase function studies) 
are emerging as particularly intriguing and timely.   
Hence, in \S\ref{irradiation}, we review the theory of these subtopics
and explore the physics and chemistry upon which they depend. 

\begin{quote}
\begin{flushleft}
{\it Brighter art thou than flaming Jupiter,\\
     When he appeared to hapless Semele;}\\
         Christopher Marlowe - The Tragical History of Doctor Faustus, Act 5, Sc. 1
\end{flushleft}
\end{quote}

\section{Irradiation, Transits, and Spectra of Close-in EGPs}
\label{irradiation}

Close-in EGPs can attain temperatures and luminosities that rival those of stars near the
main-sequence edge, despite the fact that the latter can be $\sim$50--100
times more massive (Table \ref{beast}; Guillot \etal 1996; Burrows \etal 2000).
Although the brown dwarf Gliese 229B is 10--50 times as massive as the EGP 51 Peg b, they have similar effective temperatures.
Moreover, stellar irradiation can swell the radii of such short-period 
gas giants by 20\% to 80\% (Guillot \etal 1996), thereby enhancing the magnitude
and probability of the photometric dip during a planetary transit ({\it e.g.}, HD209458b: Charbonneau \etal 1999, Henry \etal 2000;
Mazeh \etal 2000, Jha \etal 2000, Brown \etal 2000).

Hence, the most interesting, unexpected, and problematic subclass of EGPs are those found within
$\sim$0.1 A.U. of their primaries, 50--100 times closer than Jupiter is to our Sun.  At such orbital distances,
due to stellar irradiation alone, an EGP can have an effective temperature
greater than 600 K.  Indeed, the EGPs HD187123b, HD209458b,
$\tau$ Boo b, HD75289b, 51 Peg b, $\upsilon$ And b, and HD217107b
(Mayor and Queloz 1995; Marcy and Butler 1995; Butler \etal 1997,1998;
Fischer \etal 1999; Henry \etal 2000; Charbonneau \etal 2000) likely all have \teffs\ above 1000 K (see Table \ref{beast}).
This is to be compared with \teffs\ for Jupiter and Saturn of 125 K and 95 K, respectively.
Despite such proximity, these planets are stable to tidal stripping and
significant evaporation (Guillot \etal 1996).  

While direct detection and imaging of a planetary point source under the glare
of the primary star may be a few years off, there are in principle ground-based means 
other than radial-velocity techniques by which such close-in
EGPs can be detected.  They include 
astrometric techniques (Horner \etal 1998), nulling interferometry
(Hinz \etal 1998), spectral deconvolution (Charbonneau, Jha, and Noyes 1998),
and looking for the wobble in the light
centroid for a well-chosen spectral band for which the planet and star have different
absorption/emission features.  
Furthermore, proposed space telescopes/instruments, such as NGST (Next Generation
Space Telescope; Mather 2000), SIM (Space Interferometry Mission; Catanzarite \etal 1999), Eclipse (Trauger \etal 2000), 
Eddington (Penny, Favata, and Deeg 2000), Kepler (Koch \etal 2000), 
FAME (Horner \etal 1998), COROT (Michel \etal 2000), MONS (Kjeldsen, Bedding, and Christensen-Dalsgaard 2000),
and MOST (Matthews \etal 2000), may well be able to  
detect and characterize a subset of EGP systems.

Using spectral deconvolution,  
Charbonneau \etal (1999) and Cameron \etal (2000) were able to constrain
the geometric albedo of the ``roaster,'' $\tau$ Boo b, to be below
$\sim$0.3 ($\sim$0.48 \mic)  and $\sim$0.22 (0.4 \mic to 0.6 \mic), respectively, and have shown that such techniques even today
are tantalizingly close to the sensitivity required for direct detection.  
However, in perhaps one out of ten of the close-in EGPs we expect to observe planetary transits (probability $\sim$R$_{*}$/$a$).
For a giant planet, the photometric dip ($\sim$R$_{\rm p}^2$/R$_{*}^2$) in the stellar light due to a transit 
is not the $\sim$0.01\% expected for Earth-like planets,
but, due to its greater diameter, is one hundred times larger.  This is quite easily measured from the ground.
The recent discovery that one of the close-in EGPs, HD209458b, does indeed transit its
star has provided a first-of-its-kind measurement of an EGP's radius and mass
and a glimpse at what one will be able to learn once a family of these systems 
is found. In this section, we review the basics of transits, reflection spectra, and albedos,
in anticipation of the increasing relevance of such quantities to EGP studies 
in the next few years. 

\begin{quote}
\begin{flushleft}  
{\it Eppur si muove.\\
     (But it does move.)}\\
         Galileo Galilei - Apocryphal words when before the Inquisition
\end{flushleft}
\end{quote}

\subsection{The Transit of HD209458}
\label{HD209}

The HD209458b transits were the first stellar transits by
a planet with an atmosphere to be observed in over a century.  In fact,
prior to these observations, only five such events have been recorded
in human history: the transits of Venus in 1639, 1761, 1769, 1874, and 1882.
The transits of the F8V/G0V star HD209458 (at a distance of 47 parsecs) by HD209458b last
$\sim$3 hours (out of a total period of 3.524 days) and have a depth
of $\sim$1.5-2.0\%.  The ingress and egress phases each last
$\sim$25 minutes (Charbonneau \etal 2000; Henry \etal 2000). The properties of
the planet, in particular its orbital distance and radius,
scale with the properties of the star and values for the planet's radius (\rp)
around $\sim$1.4 \rj have been quoted (Henry \etal 2000; Charbonneau \etal 2000;
Mazeh \etal 2000; Jha \etal 2000), with the most precise
value, $1.347 \pm 0.060$\rj, derived from HST/STIS photometry (Brown \etal 2000).

With the availablility of such a precise photometrically-derived radius, 
the question naturally arises: what physical processes
determine the transit radius and to what pressure level does
it pertain?  Figure \ref{RMP_H3.ps} shows results of a calculation carried out
by Hubbard \etal (2000), with atmospheric radius adjusted to
provide a good match to the Brown \etal lightcurve over the wavelength
band that they used for the observation (0.582 \mic to 0.639 $\mu$m). Several
physical processes in an EGP's atmosphere are potentially important for
determining the transit radius: Rayleigh scattering, refraction, 
molecular absorption, and cloud scattering.  Hubbard \etal find that molecular
absorption dominates all other mechanisms, at least in the wavelength
interval given. In agreement with the predictions of
Seager and Sasselov (1998), due to variations in 
the molecular absorption cross sections with wavelength,
the transit radius is a function of wavelength.
In the region which determines the HST/STIS transit radius, the 
atmosphere is estimated to have a temperature of $\sim$1000 K
and where the slant optical depth is unity the pressure 
is $\sim$10 millibars.
The proper radius to compare with Jupiter is the radius corresponding to
1 bar pressure, which for Jupiter is determined to be $71492 \pm 4$ km at the
equator (Lindal \etal 1981).  The best-fit
model of Hubbard \etal (2000) has a radius at 1 bar equal to
94430 km, or 1.32 \rj. The Brown \etal (2000) error on the  
transit radius for an equivalent opaque occulting disk
of $\pm 0.06$ \rj is still more
than twice the uncertainty due to details of the atmospheric
structure (the atmospheric scale height in HD209458b might be
$\sim$0.01 \rj ).  However, with improved measurements of
the transit lightcurve due to repeated observations,
atmospheric details will ultimately play a
role in the interpretation of the data.

Such large radii for close-in EGPs were predicted by 
Guillot \etal (1996).  Importantly, since HD209458b transits its 
primary, astronomers can derive $\sin (i)$ ($i \sim 86.7^{\circ}$),
from which the planet's mass ($\sim$0.69 \mj) can be directly determined.
The ability to pin down more than one structural parameter for a given EGP
is a milestone in the emerging study of extrasolar planets and a harbinger of
what can be expected as the list of known close-in EGPs grows.

As Zapolsky and Salpeter (1969) demonstrated for planets made
of high-$Z$ material, any planet with a cold radius larger
than $\sim$0.75 \rj, must be made predominantly of hydrogen.  Using the ANEOS equation of state tables (Thompson 1990),
Burrows \etal (2000) derive that an ``olivine'' (rock) or H$_2$O (ice) planet with a mass of 0.69 \mj has a radius
of 0.31 \rj or 0.45 \rj, respectively.  Importantly, these radii are 3--4 times smaller than observed for HD209458b
and prove that HD209458b must be a hydrogen-rich gas giant; it cannot be a giant terrestrial planet or
an ice giant such as Neptune or Uranus.  This point is made graphically in Fig. \ref{guil.rock} 
from Guillot \etal (1996) which depicts the radius versus mass of an EGP for different compositions
and at 51 Peg A's age ($\sim$8 Gyr). 

Burrows \etal (2000) also found that HD209458b's large radius
is not due to mere thermal expansion of its atmosphere ($\sim$1\%), but is due to the high residual entropy
that remains throughout its bulk by dint of its early proximity to the luminous primary.
The large stellar flux does not inflate the planet, but retards
its otherwise inexorable contraction from a more extended configuration at birth.
Essentially, irradiation flattens the temperature profile at
the top of the convective zone, while at the same time moving the radiative/convective boundary inward.
Since radiative fluxes are governed by the product of thermal diffusivities and temperature
gradients and since the thermal diffusivity decreases with increasing pressure ($\propto \sim P^{-2}$),
the flux of energy out of the convective core and the rate of core entropy change are
reduced.  The upshot is a retardation of the contraction of the
planet. High stellar fluxes on a close-in EGP can have profound structural consequences.
In particular, stellar insolation can be responsible for maintaining the planet's radius
at a value 20\% to 80\% larger than that of Jupiter itself (Guillot \etal 1996; Burrows \etal 2000).
Figure \ref{transit.rad} depicts two possible \rp{--$t$} trajectories for HD209458b, if born and fixed at 0.045 A.U.
A box of empirical ages and \rp{'s} (Mazeh \etal 2000) is superposed.
As the figure demonstrates, one can fit the mass-radius-age combination 
of HD209458b, though ambiguities in the Bond albedo, the degree of redistribution
of heat to the night side, and thermal
conductivities do not yet allow one to do so with precision.
Also included on Fig. \ref{transit.rad} is a theoretical evolutionary trajectory for a 0.69-\mj EGP in isolation.

As can be seen in Fig. \ref{transit.rad}, there is a great difference between the theoretical radius
of an isolated and of an irradiated EGP.  Since the scale-height effect is miniscule,
the \rp{--$t$} trajectory of the isolated EGP (model I) immediately suggests that if HD209458b were
allowed to dwell at large orbital distances ($\geq$0.5 A.U.) for more than a 
few$\times 10^7$ years, its observed radius could not be reproduced.
It is at such ages that the radius of an isolated 0.69-\mj EGP falls below
HD209458b's observed radius.  This implies either that such a planet was formed near its current orbital distance or that
it migrated in from larger distances ($\geq$0.5 A.U.), no later than a few times $10^7$ years of birth.
This is the first derived constraint on the history of an EGP and suggests the potential
power of coupled transit and evolutionary studies.

\subsection{EGP Albedos and Reflection Spectra}
\label{albedo}

Our discussion of albedos follows closely that in 
Sudarsky, Burrows, and Pinto (2000, hereafter SBP),
to which the reader is referred for further details.
The albedo of an object is simply the fraction of light that the object reflects.
However, there are several different types of albedos.  The {\it geometric\/} albedo, $A_g$,
refers to the reflectivity of the object at full phase ($\alpha = 0$, where $\alpha$
represents the object's phase angle) relative to that
by a perfect Lambert disk of the same radius under the same incident flux.
The {\it spherical\/} albedo, $A_s$,
refers to the fraction of incident light reflected by a sphere at all angles.  Usually stated as a function
of wavelength, it is obtained by integrating the reflected flux over all phase angles.
The {\it Bond\/} albedo, $A_B$, is the ratio of the total reflected and total incident
powers.  It is obtained by weighting the spherical albedo by the spectrum of the
illuminating source and integrating over all wavelengths.  This is the quantity with
which the equilibrium temperature of an irradiated planet is determined.  Ignoring
internal luminosity and assuming that the planet reradiates equally into $4\pi$ steradians, 
the equilibrium temperature is equal to $(R_*/2a)^{1/2}{\rm T_{*}}(1-A_B)^{1/4}$ (Table \ref{beast}),
where T$_*$ and $R_*$ are the effective temperature and radius, respectively, of the primary star.

Spherical, geometric, and Bond albedos of objects are strong functions of their compositions.
Within the solar system, they vary substantially with wavelength,
and from object to object.  At short wavelengths, 
gaseous atmospheres can have high albedos due to
Rayleigh scattering and low albedos at longer wavelengths
due to molecular ro-vibrational absorption.  Due to water 
and methane absorption, the latter should be the case 
for cloudless EGPs.  Icy condensates, whether they reside
on a surface or are present in an upper atmosphere, are highly reflective
and increase the albedo.  Other condensates, such as silicates or non-equilibrium
products of photolysis, can lower the albedo,
but longward of $\sim\! 0.6$ \mic most clouds at 
altitude raise albedos substantially.

The theoretical study of EGP albedos and reflection spectra is still largely in its infancy.
Marley \etal (1999) explored a range of EGP geometric
and Bond albedos using temperature-pressure profiles of EGPs in isolation
({\it i.e.}, no stellar insolation), while
Goukenleuque \etal (2000) modeled 51 Peg b in radiative equilibrium
and Seager and Sasselov (1998) explored radiative-convective models
of EGPs under strong stellar insolation.
The theoretical reflection spectra generated by Goukenleuque 
\etal (2000) and by Seager, Whitney, and Sasselov (2000) 
are particularly informative.  Figure \ref{seager} from the latter depicts a full-phase spectrum
of the close-in EGP 51 Peg b.   The water, methane, and alkali metal absorption 
features are clearly in evidence and will be primary diagnostics
of this planet when it is finally directly detected.   

The ratio of the planet's flux to the star's flux is given by:

\begin{equation}
\frac{{\cal{F}}_p}{{\cal{F}}_*} = A_g\Bigl(\frac{{\rm R_p}}{a}\Bigr)^2 P(\alpha),
\label{lambert}
\end{equation}
where $A_g$ is the geometric albedo, $a$ is the planet's semi-major axis, 
$\alpha$ is given by the expression $\cos\alpha = - \sin i\sin(2\pi\Phi)$, $i$ is the
orbital inclination angle, $\Phi$ is the orbital phase (from 0 to 1),
and $P(\alpha)$ is the phase function normalized to 1 at full phase ($\alpha = 0$; $\Phi = -0.25$).  
For a Lambert sphere, $A_g = 2/3$ and for Jupiter at $\lambda = 4800$\AA\ $A_g \sim 0.46$.
The most important feature of eq. (\ref{lambert}) is the geometric factor $(R_p/a)^2$, which for 51 Peg b is $\sim 10^{-4}$.
Seager, Whitney, and Sasselov (2000) explored the angular phase functions
of EGPs in  reflection and demonstrated that, due in part to forward scattering
by large cloud particles, the range of phase angles for which the reflection
is high can be significantly more narrow than what the simple Lambert phase function 
($P(\alpha) = (\sin\alpha + (\pi - \alpha)\cos\alpha)/\pi$)
gives. Hence, in the early stages of direct detection campaigns, one does not 
want to observe far away from full phase.  

SBP recently 
attempted to establish a general understanding
of the systematics of the albedo and reflection spectra of EGPs, tying them to
their overall compositions and clouds.
For those preliminary calculations, SBP defined five representative
composition classes based loosely on T$_{\rm{eff}}$.
The classification of EGPs into five composition classes,
related to T$_{\rm{eff}}$, is instructive,
since it can be shown that the albedos of objects within each of these classes
exhibit similar features and values.
The ``Jovian'' Class I objects (T$_{\rm{eff}} \sles$ 150 K)
are characterized by the presence of ammonia clouds and gaseous methane.
In somewhat warmer objects (T$_{\rm{eff}}$
$\ge$ 250 K), ammonia is in its
gaseous state, but the upper troposphere contains condensed H$_2$O.
These objects are designated Class II, or ``water cloud,'' EGPs and also contain
a large abundance of gaseous methane.
Class III, or ``clear'' EGPs, are so named because they are
too hot (T$_{\rm{eff}} \sgreat$ 500 K)
for significant H$_2$O condensation and so are not expected to have
clouds in their atmospheres.  
Absorption by gaseous water, methane, molecular hydrogen (via CIA), and neutral alkali metals, together with
the absence of dominating cloud layers, give Class III EGPs the lowest Bond and geometric albedos
of any class. In hotter EGPs (900 K $\sles$ T$_{\rm{eff}}$ $\sles$ 1500 K; Class IV)
the troposphere is expected to contain significant abundances
of absorbing neutral sodium and potassium gases above a silicate cloud layer and the albedo will be low.  However, the hottest
(T$_{\rm{eff}} \sgreat$ 1500) and/or least massive EGPs with low gravities (g $\sles 10^3$ cm s$^{-2}$)
have a silicate layer located so high in the atmosphere that much of
the incoming radiation is reflected back out into space before being absorbed by 
neutral alkali metals or molecules (Seager and Sasselov 1998).
SBP designated these highly reflective EGPs Class V.
Hence, SBP concluded that
neither the Bond nor the geometric albedos of EGPs are monotonic with T$_{\rm{eff}}$.  For instance, around a G2V star,
the Bond albedos are $\sim$0.5, $\sim$0.8, $\sim$0.1,  $\sim$0.03, and $\sim$0.55 for Classes I through V.
van de Hulst (1974) has derived a useful analytic expression for the spherical albedo of
a uniform, homogeneous atmosphere with a scattering albedo, $\omega$ ($= \sigma_{\rm scat}/(\sigma_{\rm scat} +  \sigma_{\rm abs})$), 
and a scattering asymmetry factor, $g = <\cos\theta>$: 

\begin{equation}
A_S = \frac{\Bigl(1 -0.139s\Bigr)\Bigl(1-s\Bigr)}{\Bigl(1 + 1.17s\Bigr)},
\end{equation}
where $s = \bigl((1-\omega)/(1-g\omega)\bigr)^{1/2}$.  Though this formula can't handle stratification,
it summarizes the basic influence of absorption and of scattering asymmetry
on albedos.  The appropriate integration over wavelength yields the corresponding Bond albedo.

Direct photometric and spectroscopic detection
of close-in EGPs is likely within the next few years.
The SBP set of EGP albedos serve as a useful guide to the prominent features and systematics
over a full range of EGP effective temperatures, for \teffs\ from $\sim$100 K to
1700 K.  However, full radiative/convective and evolutionary 
modeling of a given EGP at a specific orbital distance
from its central star (of given spectral type), and of specific
mass, age, and composition is necessary for a detailed understanding of a particular
object.  Furthermore, tidal effects will enforce corotation for nearby planets
and this will require an understanding of the mechanical (wind) transport of
heat to the planet's night side in order to determine the energy budget 
and temperature profile of the close-in planet's atmosphere (Showman and Guillot 2000).
However, in the spirit of preliminary approximation,
Figure \ref{refspect} depicts reflection spectra of some representative EGPs from Classes II through V.  
The strength of the various water and methane features depends upon the presence of clouds and the cloud model assumed.   
Hence, Fig. \ref{refspect} should be viewed as no more than a guide to the gross systematics of EGP reflection
spectra.  Nevertheless, the significant variety of EGP reflection spectra is evident and will be diagnostically useful.

\section{The Giant Planets of the Solar System}
\label{jovian}

The precisely known values of the age, mass, radius, intrinsic
luminosity, and chemical composition of the hydrogen-rich
giant planets, Jupiter and Saturn, allow us  
to calibrate our theoretical calculations
of the structure and evolution of EGPs and SMOs. In contrast,
the lower-mass ice giant planets, Uranus and Neptune,
(masses $\sim$0.05 \mj) plot slightly above the radius versus mass
curve for pure H$_2$O on Fig. \ref{guil.rock} and are thus inferred to contain
mostly H$_2$O and other hydrides of C, N, and O, with a
superficial layer of H$_2$ and He
(Hubbard, Podolak, and Stevenson 1995). 
Uranus and Neptune are thus so different from the hydrogen-rich
bodies that they are not currently useful as ``calibrators'' (although
this situation may change when even lower mass EGPs are detected).

In contrast, much of the same physics which governs the evolution of
EGPs and SMOs is also applicable to Jupiter and Saturn.
This point is made in Fig. \ref{TPbill}, which shows interior pressure-temperature
profiles for Jupiter and the brown dwarf Gl 229B at various ages,
superimposed on a phase diagram for dense hydrogen.
Phase boundaries for pure hydrogen are shown in green, with
dashed portions representing particularly uncertain boundaries.
At pressures below about 1 Mbar ($\sim$one million atmospheres), hydrogen is in the form of a
nearly-isentropic supercritical liquid which merges seamlessly
with the atmosphere.  The nature of the transition from liquid
molecular to liquid metallic hydrogen (terminating at critical 
point ``SCVH'') is based on the theory of
Saumon, Chabrier, and Van Horn (1995), 
although experimental data
in the vicinity of the curve (Collins \etal 1995)
show no evidence of a discontinuity.
The areas shaded in yellow and light blue pertain to a solar mixture
of H and He, and indicate the region in P-T space where such a
mixture is theoretically predicted to separate into two liquid
phases, one He-rich and the other He-poor.  The yellow region
shows the region of such immiscibility predicted by
Stevenson and Salpeter (1977), 
while the blue region marked with the letter ``P.'' represents
a prediction for the same phenomenon, calculated by
Pfaffenzeller, Hohl, and Ballone (1995).

The red dotted curves show the interior isentrope for the
representative brown dwarf at an early age (7 Myr) and at
an advanced age of 2 Gyr, nominally the present Gl 229B.
The red dash-dot curve shows the evolutionary trajectory of
its central pressure and temperature, illustrating the
phenomenon of the achievement of a maximum temperature and
its subsequent decline owing to the onset of
electron degeneracy (Fig. \ref{tc.ps}).  Since
the maximum central temperature exceeds the threshold for
deuterium fusion (shown in black), the brown dwarf Gl 229B
should be depleted in deuterium.

A similar set of curves (in blue) is shown for the evolution of
Jupiter.  The principal points that Fig. \ref{TPbill} makes are:
(a) the present P-T profile of Gl 229B is nearly coincident
with that of the early Jupiter (age $\sim$0.1 Gyr), and, thus,
this object's envelope structure reprises that of the early
Jupiter,
(b) both Jupiter and Gl 229B could potentially be affected
by a first-order phase transition between H$_2$ and metallic H
(at least, if the SCVH theory is correct), and 
(c) in the late (current) stages of evolution of Jupiter
possible H-He immiscibility could affect the structure and
luminosity of the object (if the Stevenson-Salpeter theory
is correct, but not if the Pfaffenzeller \etal theory
is correct).

While Jupiter and Saturn provide useful test cases for models
of EGPs, complications exist.  First, neither Jupiter nor
Saturn are of precisely solar composition.  A careful study
of Jupiter's composition by Guillot, Gautier, and Hubbard (1997) finds that the
metallicity (Z-component) of the latter is between one and
seven times solar.  While there is evidence for a dense core
in Jupiter, it is apparently not a large one and most of
the Z-component is distributed throughout the planet.
Evidently, the formation of Jupiter was an inefficient
process, leading to a modest enhancement of the more refractory
(and denser) compounds relative to the most volatile ones
(principally H and He).  This is a hallmark of a planetary,
as opposed to a ``stellar,'' mode of formation.

Saturn's bulk composition is even more strongly non-solar
and there is clear evidence in the gravity field for a
sizeable core, comprising perhaps 20\% of the total
planetary mass (Hubbard and Stevenson 1984).  
Although there is some evidence for He depletion in
the atmosphere, the depletion with respect to solar
composition is less extreme than initially reported,
and could even be zero (Conrath and Gautier 2000),
consistent with the
Pfaffenzeller \etal phase diagram.  Nevertheless,
there is a long-standing discrepancy between the observed
luminosity of Saturn and the prediction based on a
straightforward application of the cooling theory
for EGPs: Saturn's present luminosity is too large,
compared with the predicted value.  This point is illustrated
in Fig. \ref{pfaff}, which is taken from Hubbard \etal (1999).
The curves marked with crosses are the calculated
cooling curves for isolated planets of Jupiter's and Saturn's
mass, essentially from the EGP theory.  The curves marked with
solid and open circles are cooling curves calculated 
with allowance for atmospheric heating from solar luminosity,
assuming variable and constant solar luminosity, respectively.
All of the smooth curves are extrapolated to the present
epoch (dashed lines) assuming homogeneous interior composition,
while the abrupt change in slope near the present \teff
shows the effect of interior differentiation into dense
and less-dense components under various assumptions.  The
horizontal error bars show alternative extensions of the
cooling age for assumed differentiation.
Clearly differentiation, which is not included in the standard
EGP theory, is needed to explain Saturn's evolution.  In view
of the questions about whether this is consistent with
the H-He phase diagram, perhaps some other abundant dense
component is presently differentiating in Saturn.   
If true, this may be one means by which to distinguish the origin, ``planetary''
or ``stellar,'' of an old, sub-jovian mass SMO.
However, it is reassuring that the EGP theory works well for the
more massive (and slightly hotter) Jupiter.

\section{Conclusions}
\label{conclusions}

A major theme of this review is that, from the 
theoretical perspective, however different their modes of formation,
EGPs and brown dwarfs are essentially the same. 
Whatever their masses and metallicities, and 
whether or not one class nucleates at formation around an 
ice/rock core, each subclass of hydrogen-rich object can be 
treated with the same basic theoretical tools.  
As such, in a very useful sense, the chemistry, spectra, 
atmospheric physics, opacities, equation of state, 
and nuclear physics that we explored in this paper  
apply universally across the SMO continuum,
even to EGPs in close proximity to their central stars.    
Such universality is at the very heart of Figs. \ref{lum.ps} and \ref{teff}.
The resulting science is a fruitful merger of planetary and classical astronomy
(drawing from the perspectives of each) 
that is more than the sum of its parts.
 
We have attempted here to summarize a still emerging theory 
for objects below the main sequence edge that encompasses
three orders of magnitude in both mass and age
and a factor of $\sim$30 in \teff.
Though there has been incredible 
progress in SMO studies since 
our earlier review (Burrows and Liebert 1993), much
concerning their spectra, compositions, band and line strengths, 
cloud properties, statistics, and atmospheric physics remains to be determined,
understood, and explained.  Fortunately, as with 
any healthy science, the theory is now being driven    
by an avalanche of new measurements and observations. 

The prospects for additional discoveries of extrasolar planets and brown
dwarfs, and their ever-more detailed characterization, are excellent. For EGPs,
continued radial velocity discoveries in the inner few A.U. around nearby stars
will be extended with more sensitive surveys around M dwarfs.  
In perhaps 1\% of systems, we can expect observations of
transits of EGPs across their parent stars, as well as  
(for many more systems) astrometric detections from 
the ground (with the Keck, VLT, and Large Binocular Telescope (LBT) interferometers)
and from space (with FAME and SIM).  Several
spaceborne efforts to detect the transits of planets as small as Earth (with COROT, MOST, MONS)
or to directly detect EGP emissions/reflections (for instance, with the LBT, NGSS, Eclipse, Eddington) are proposed,
planned, or imminent.  Likewise, microlensing projects to survey larger volumes of space
for brown dwarfs, EGPs, and even Earths are underway on the ground and
planned for space.  Ultimately, large spaceborne interferometers such as that
conceived for the Terrestrial Planet Finder (TPF) mission, will not only image planetary 
systems, but will also provide moderate-resolution spectra of
giant and terrestral planets orbiting nearby stars.

Importantly, the photometric bandpasses of the IRAC (3.6 \mic, 4.5 \mic, 5.8 \mic
and 8.0 \mic) and the MIPS ($\ge$20 \mic) cameras on SIRTF (the Space 
InfraRed Telescope Facility) include those optimal for detection of brown
dwarfs cooler than can be found by the current 2MASS, SDSS and DENIS
surveys from the ground.  Figure \ref{15MJ-30} depicts theoretical spectra
from 1.0 \mic to 30 \mic of a 15 \mj brown dwarf at various epochs.  Superposed
are approximate sensitivities of SIRTF, NGST, TPF, and Keck.  Particularly around 5 \mic,
but also longward of 10 \mic, theory predicts that cool brown dwarfs 
will be a major activity of space-borne telescopes in the next decade.
Hence, while these telescopes will be able to survey only a modest
area of the sky, they will be excellent for detecting brown dwarfs in the
mid-infrared, for targeting known
field L and T dwarfs for companions of lower mass, and for 
discovering field ``water cloud'' dwarfs (\S\ref{spectra} and \S\ref{albedo})
with \teffs below those of the T dwarfs.

Given the increasing
area of mirror glass that is being coupled to sensitive detectors, 
optical and infrared spectra of L and T dwarfs, as well as of a handful of
cooler giant planets, will be routinely obtained from the ground.  Moderate 
class telescopes (2- to 4-meter diameter), run by consortia of universities, are likely to proliferate 
and these facilities are excellent for completing L-dwarf, T-dwarf, and water-cloud-dwarf surveys in the solar
neighborhood.  As the oversubscribed 10-meter Kecks are 
joined by the 6.5-meter MMT and Magellan telescopes (now being
commissioned), the 8-meter Subaru, the twin-8.4 meter LBT,
the 8-meter Geminis, the 9-meter Hobby-Eberly Telescope, and the European VLT (four 8-meter mirrors) (among others),
deeper searches, and higher resolution spectra, of nearby T dwarfs will become possible and routine.
Ultimately, in order to spectroscopically investigate L and T dwarfs
to the level that is now possible for Jupiter in our own solar system will require a
new generation of telescopes with apertures of 30 meters or more;
unbelievably, these are now in the planning stages.

The study of brown dwarfs and EGPs is very much germane to understanding the
occurrence and properties of planetary systems, the baryonic content and
chemical evolution of the cosmos, and the relationship (in both genesis and
physical properties) between stars and objects not massive enough to ever
become stars.  By detecting and characterizing brown dwarfs and EGPs, we
extend our knowledge of the cosmos from the ubiquitous macroscale of stars
ever closer to the instinctively appealing, and human, scale of worlds like our
own.

\section*{Acknowledgments}

The authors are pleased to thank Mark Marley, Didier Saumon, Tristan Guillot,
David Sudarsky, Isabelle Baraffe, Gilles Chabrier, 
France Allard, Christopher Sharp, Richard Freedman, the Ben Oppenheimer(s),
Jonathan Fortney, Curtis Cooper, Davy Kirkpatrick, Adam Burgasser, Neill Reid, 
Sandy Leggett, and Sara Seager for stimulating collaboration,
conversations, encouragement, insights, and/or figure permissions. 
We would also like to thank Franca D'Antona and Davy Kirkpatrick for carefully reading the
manuscript.  Our theoretical understanding ripened and the new subject
of substellar-mass objects was born with the aid of NASA's financial support
(NAG5-7073, NAG5-7499, NAG5-711), here gratefully acknowledged. 



\begin{center}
\begin{table}
\caption{The Bestiary of EGPs: A Compendium of all the radial-velocity ``planet" discoveries circa August 2000,
in order of increasing orbital semi-major axis. Also shown are Jupiter and Saturn.
Included in this ``bestiary" are the object name, star mass, luminosity, distance, stellar metallicity,
$m_p sin(i)$ (M), orbital semi-major axis, period, eccentricity, an estimate of the planet's \teff, and
an approximate stellar age. See \S\ref{basics:egp} in text for references.}
\label{beast}
\begin{tabular}{llllllllllll}

\undertext{Object}&\undertext{Star}&\undertext{M$_\star$ (M$_\odot$)}&\undertext{L$_\star$ (L$_\odot$)}&\undertext{d (pc)}&\undertext{[Fe/H]}&\undertext{M (M$_J$)}&\undertext{$a$ (AU)}&\undertext{P (days)}&\undertext{$e$}&\undertext{T$_{eff}$ (K)}&\undertext{Age} \\

HD83443b&K0V&0.79&0.93&43.54&+0.33&$\sgreat$0.35&0.038&2.986&0.079&1200&\\
HD46375&K1IV&1.0&$\sim$1&33.4&+0.34&$\sgreat$0.24&0.041&3.024&0.04&1250& $>$5\\
HD187123&G3V&1.0&1.35&48&+0.16&$\sgreat$0.52&0.0415&3.097&0.03&1460&?\\
HD209458&F8V&1.1&2.0&47&+0.0&$\sim$0.69&0.045&3.52&0.0&1270&4.5\\
$\tau$ Boo b&F7V&1.42&3.2&17&+0.27&$\sgreat$4.14&0.046&3.313&0.0162&1600&1 Gyr\\
BD-103166&&&&&&$\sgreat$0.48&0.046&3.487&0.05&&\\
HD75289&G0V&1.05&1.99&29&+0.29&$\sgreat$0.46&0.048&3.51&0.0&1390&5 Gyr\\
51 Peg b&G2.5V&1.0&1.0&15.4&+0.21&$\sgreat$0.45&0.05&4.23&0.0&1240&8\\
HD98230&G0V&1.1&1.5&7.3&-0.12&$\sgreat$37&0.05&3.98&0.0&?&?\\
$\upsilon$ And b&F7V&1.25&2.5&17.6&+0.17&$\sgreat$0.71&0.059&4.617&0.034&1430&3\\
HD168746&G5V&0.92&$\sim$1.1&43.1&&$\sgreat$0.24&0.066&6.409&0.0&1000& \\
HD217107&G7V&0.96&1.0&37&+0.29&$\sgreat$1.28&0.07&7.11&0.14&1030&7.8\\
HD162020&K2V&0.7&0.24&31.26&&$\sgreat$13.7&0.072&8.43&0.284&700& \\
HD130322&K0V&0.79&0.5&33.6&$\sim$0.0&$\sgreat$1.08&0.08&10.7&0.06&810&? \\
HD108147&G0V&1.05&$\sim$1.93&38.57&&$\sgreat$0.34&0.098&10.88&0.56&800&$\sim$2 \\
55 Cnc b&G8V&0.85&0.5&13.4&+0.29&$\sgreat$0.84&0.11&14.76&0.051&690&5\\
GJ 86 Ab&K1V&0.79&0.4&11&-0.3&$\sgreat$4.9&0.11&15.83&0.05&660&? \\
HD38529&G4IV&1.4&&42.4&+0.23&$\sgreat$0.77&0.13&14.32&0.27&&\\
HD195019&G3V&0.98&1.0&20&$\sim$0.0&$\sgreat$3.4&0.14&18.3&0.05&720&3\\
HD6434&G3IV&1.0&1.12&40.32&&$\sgreat$0.48&0.15&22.1&0.30&650&3.7\\
HD192263&K2V&0.75&0.34&19.9&-0.20&$\sgreat$0.78&0.15&24.36&0.22&540&?\\
HD83443c&K0V&0.79&0.93&43.54&+0.33&$\sgreat$0.15&0.174&29.8&0.42&&\\
GJ 876 b&M4V&0.32&0.01&4.72&&$\sgreat$2.45&0.2&60.8&0.24&180&\\
$\rho$ CrB b&G0V&1.1&1.77&17.4&-0.19&$\sgreat$1.13&0.23&39.65&0.028&670&10\\
HR7875 b&F8V&1.2&2.1&25&-0.46&$\sgreat$0.69&0.25&42.5&0.429&650&\\
HD168443b&G8IV&0.95&2.1&33&-0.14&$\sgreat$7.37&0.29&58.14&0.52&620&7-10\\
HD121504&G2V&1.0&1.58&44.37&+0.3&$\sgreat$0.89&0.32&64.6&0.13&500&2.8\\
HD16141&G5IV&1.0&$\sim$1&35.9&+0.02&$\sgreat$0.22&0.35&75.8&0.28&420& $>$5\\
HD114762&F9V&1.15&1.8&28&-0.60&$\sgreat$10&0.38&84&0.25&510&9.4\\
70 Vir b&G4V&0.95&0.8&18.1&-0.03&$\sgreat$6.9&0.45&116.7&0.40&380&9\\
HD52265&G0V&1.05&1.98&28.07&&$\sgreat$1.07&0.48&119&0.38&400&4 \\
HD1237&&&&&&$\sgreat$3.45&0.505&133.8&0.51&&\\
HD37124&G4V&0.91&&33&-0.32&$\sgreat$1.13&0.55&155&0.31&350&\\
HD202206&G6V&0.9&1.12&46.34&&$\sgreat$14.7&0.77&259&0.422&300&0.6 \\
HD12661&&&&&&$\sgreat$2.83&0.799&250.2&0.20&&\\
HD134987&G5V&1.05&1.34&26&+0.23&$\sgreat$1.58&0.81&260&0.24&320&\\
HD169830&F8V&1.4&4.63&36.32&&$\sgreat$2.96&0.82&230.4&0.34&350&4 \\
$\upsilon$ And c&F7V&1.25&2.5&17.6&+0.17&$\sgreat$2.11&0.83&241.2&0.18&370&3\\
HD89744&&&&&&$\sgreat$7.17&0.883&256&0.70&&\\
HD92788&G5V&0.95&1.1&32.32&+0.25&$\sgreat$3.8&0.94&340&0.36&250&\\
$\iota$ Hor&&&&&&$\sgreat$2.98&0.97&320&0.16&&\\
HD177830&K2IV&1.15&&59&&$\sgreat$1.22&1.1&391&0.41&380&\\
HR5568 b&K4V&0.71&0.13&6&$\sim$0.0&$\sgreat$0.75&1.0&400&&160?&\\
HD210277&G7V&0.92&0.93&21&+0.24&$\sgreat$1.28&1.15&437&0.45&180&7-10\\
HD82943&G0V&1.05&1.54&27.46&&$\sgreat$2.2&1.16&443&0.61&300&5 \\
HR810 b&G0V&1.1&1.5&15.5&&$\sgreat$2.0&1.2&599.4&0.492&190&\\
HD19994&F8V&1.35&3.8&22.38&+0.23&$\sgreat$1.8&1.3&454&0.2&300&3\\
HD222582&G3V&1.0&1.2&42&-0.01&$\sgreat$5.29&1.35&576&0.71&250&\\
16 Cyg Bb&G2.5V&1.0&1.0?&22&+0.11&$\sgreat$1.66&1.7&2.19 yrs&0.68&160&5?\\
47 UMa b&G0V&1.1&1.5&14.1&+0.01&$\sgreat$2.5&2.1&2.98 yrs&0.03&160&7\\
HD10697&G5IV&1.1&&32.6&+0.15&$38\pm 13$&2.12&1072.3&0.12&280&\\
HD190228&G5IV&1.3&4.38&62.1&&$\sgreat$4.99&2.31&1127.5&0.43&&\\
$\upsilon$ And d&F7V&1.25&2.5&17.6&+0.17&$\sgreat$4.61&2.5&3.47 yrs&0.41&230&3\\
HD168443c&G8IV&0.95&2.1&33&-0.14&$\sgreat$16.1&2.7&1660&&&7-10\\
14 Her b&K0V&0.82&0.42&18&+0.50&$\sgreat$5.44&2.84&4.4 yrs&0.37&170?&\\  
55 Cnc c&G8V&0.85&0.5&13.4&+0.29&$\sgreat$3.14&3.8&12 yrs&0.24&200&5\\
Jupiter&G2V&1.0&1.0&0.0&0.0&1.00&5.2&11.86 yrs&0.048&125&4.6\\
Saturn&G2V&1.0&1.0&0.0&0.0&0.3&9.54&29.46 yrs&0.056&95&4.6\\

\end{tabular}
\end{table}
\end{center}

\begin{table} 
\caption{Anders and Grevesse (1989) solar abundances by number.}
\label{anders}
\begin{tabular}{l|c||l|c}
Element                &Abundance    &Element   &Abundance    \\
\hline
H   &  $9.10\times 10^{-1}$  &  Ni  &  $1.61\times 10^{-6}$     \\
He  &  $8.87\times 10^{-2}$  &  Cr  &  $4.40\times 10^{-7}$     \\
O   &  $7.76\times 10^{-4}$  &  P   &  $3.39\times 10^{-7}$     \\
C   &  $3.29\times 10^{-4}$  &  Mn  &  $3.11\times 10^{-7}$     \\
Ne  &  $1.12\times 10^{-4}$  &  Cl  &  $1.71\times 10^{-7}$     \\
N   &  $1.02\times 10^{-4}$  &  K   &  $1.23\times 10^{-7}$     \\
Mg  &  $3.49\times 10^{-5}$  &  Ti  &  $7.83\times 10^{-8}$     \\
Si  &  $3.26\times 10^{-5}$  &  Co  &  $7.34\times 10^{-8}$     \\
Fe  &  $2.94\times 10^{-5}$  &  F   &  $2.75\times 10^{-8}$     \\
S   &  $1.68\times 10^{-5}$  &  V   &  $9.56\times 10^{-9}$     \\
Ar  &  $3.29\times 10^{-6}$  &  Li  &  $1.86\times 10^{-9}$     \\
Al  &  $2.77\times 10^{-6}$  &  Rb  &  $2.31\times 10^{-10}$    \\
Ca  &  $1.99\times 10^{-6}$  &  Cs  &  $1.21\times 10^{-11}$    \\
Na  &  $1.87\times 10^{-6}$ \\
\end{tabular}
\end{table}

\begin{table}
\caption{Known L Dwarf or Brown Dwarf Binaries, taken from Reid \etal (2000b).
$\Delta$ is the binary separation in A.U. and $q$ is the mass ratio.
M$_{pri}$ is the mass of the primary and M$_{sec}$ is the mass of the secondary.
See text in \S\ref{pop} for a discussion. \\}
\label{companions}
\begin{center}
\begin{tabular}{lcccc}
System & M$_{pri}$ ($M_\odot$)& M$_{sec}$ ($M_\odot$) & $q^1$ & $\Delta$ AU \\
\tableline
PPl 15$^2$ &  0.07 & 0.06 & 0.86 & 0.03  \\
HD 10697 & 1.10 & 0.04 & 0.035 & 0.07  \\
2M0746 & $>$0.06 & $>0.06$ & 1.0 & 2.7  \\
2M0920 & 0.06-0.075 & 0.06-0.075 & 0.95 & 3.2  \\
2M0850 & $<0.06$ & $<0.06$ & 0.75 & 4.4  \\
DENIS 1228&  $<0.06$ & $<0.06$ & $\sim 1$ & 4.9  \\
2M1146 & $<0.06$ & $<0.06$ & $\sim 1$ & 7.6  \\
DENIS 0205 & 0.06-0.09 & 0.06-0.09 &  $\sim 1$ & 9.2  \\
Gl 229B & 0.5 & $\sim0.045$ & $\sim 0.1$ & 44  \\
TWA 5$^{2,3}$ & 0.4 & 0.025 & 0.06 & 100 \\
GD 165B & $> 1$ & $<0.08$ & $<.08$& 110  \\
HR 7329B & $\sim5$ & $<0.05$ & $<.01$ & 200  \\
GJ 1048B & $\sim0.7$ & $<0.08$ & $<0.11$ & 250  \\
G196-3B & 0.5 & $\sim0.025$ & $\sim 0.05$ & 340  \\
GJ 1001B & 0.4 & $\sim0.05$ & $\sim 0.13$ & 180 \\
Gl 570D & 0.7 & $\sim0.05$ & $\sim 0.07$ & 1525  \\
Gl 417B & 1.0  & $\sim0.035$& $\sim 0.035$ & 2000  \\
Gl 584C & 1.0 & $\sim0.060$& $\sim 0.060$ & 3600 \\
\end{tabular}
\end{center}
$^1$ Mass ratios for L dwarf/L dwarf systems are based on the relative K-band luminosity

$^2$ Members of Pleiades cluster or TW Hydrae association

$^3$ High resolution spectroscopy indicates that several other stars in this
moving group are binary or multiple systems.
\end{table}

\begin{table}
\caption{Gliese 229B was the first unimpeachable brown dwarf (T$_{\rm eff}\sim 950$K),
discovered by Oppenheimer \etal (1995). $^{a}$
The first 2MASS T dwarf discoveries (Burgasser \etal 1999). $^{b}$ Gliese 570D orbiting a double M dwarf system;
perhaps the coolest ($\sim$700-750 K) known T dwarf (Burgasser \etal 2000). $^{c}$ A. Burgasser (private communication). $^{d}$
First Sloan T dwarf (Strauss \etal 1999); Shows KI (7700 \AA) and NaI (5890 \AA) absorption
features (Liebert \etal 2000). $^{e}$ Tsvetanov \etal 1999.  $^{f}$ putative missing links between
the L and T dwarfs (Leggett \etal 2000a). $^{g}$ discovered and characterized by NTT/VLT; $\sim$ 90 parsecs distant (Cuby \etal 1999).
}
\label{adam}
\begin{tabular}{cccc}
 \undertext{2MASS}$^{a}$&\undertext{2MASS, unpub.}$^{c}$&\undertext{Sloan}&\undertext{NTT/VLT}\\
 2MASS J1047+21&2MASS J0243-24&SDSS J1624+00$^{d}$&NTTDF 1205-07$^{g}$\\
 2MASS J1217-03&2MASS J0559-14&SDSS J1346-00$^{e}$&\\ 
 2MASS J1225-27&2MASS J0727+17&SDSS J0539-00$^{f}$&\\ 
 2MASS J1237+65&2MASS J0937+29&SDSS J0837-00$^{f}$&\\ 
 2MASS J1457-21$^{b}$&&SDSS J1021-03$^{f}$&\\
 &&SDSS J1254-01$^{f}$&\\
\end{tabular}
\end{table}

\newpage


\begin{figure}
\caption{Evolution of the luminosity (in L${_\odot}$) of isolated solar--metallicity red dwarf stars and substellar-mass objects
versus age (in years).
The stars are shown in blue, those brown dwarfs above 13 \mj are shown in green, and brown dwarfs/EGPs equal to or below 13 \mj are
shown in red.  Though the color categories are based on deuterium or light hydrogen burning, they
should be considered arbitrary vis \`a vis whether the object in question is a brown dwarf or a planet, sensibly 
distinguished on the basis of origin.  
The masses of the SMOs/stars portrayed are 0.3, 0.5, 1.0, 2.0, 3.0, 4.0, 5.0, 6.0,
7.0, 8.0, 9.0, 10.0 11.0, 12.0, 13.0, and 15.0 \mj and 0.02, 0.025, 0.03, 0.035, 0.04,
0.045, 0.05, 0.055, 0.06, 0.065, 0.07, 0.075, 0.08, 0.085, 0.09, 0.095, 0.1,
0.15, and 0.2 \mo ($\equiv 211$ \mj).  For a given object, the gold dots mark when 50\%
of the deuterium has burned and the magenta dots mark when 50\%
of the lithium has burned.  Note that the lithium sequence penetrates into the brown dwarf regime
near 0.065 \mo, below the HBMM.
}
\label{lum.ps}
\end{figure}

\begin{figure}
\caption{The central temperature (T$_{\rm c}$) in Kelvin versus the logarithm (base ten) of the age (in Gyr) for the 
same mass set of SMOs presented in Fig. \ref{lum.ps}.  As in Fig. \ref{lum.ps}, the red lines are for models
with masses equal to or below 13 \mj, the green lines are for objects above 13 \mj and below the edge of the 
main sequence, and the blue are for stars (red dwarfs) up to 0.2 \mo.  See the text for a discussion
of the pertinent features.}
\label{tc.ps}
\end{figure}

\begin{figure}
\caption{The radius (in units of 10$^9$ centimeters) of SMOs with the masses given in Fig. \ref{lum.ps} 
versus the log$_{10}$ of the age (in Gyr).  The same color scheme that was used in Fig. \ref{lum.ps} is used here.
Red is for the low-mass SMOs, green is for the intermediate-mass SMOs, and blue is for the stars.
Also shown is the radius of Jupiter.  Note that the radii are not monotonic with mass
and that they cluster near the radius of Jupiter at late times, despite the wide range of masses from
0.3 \mj to 0.2 \mo represented.  See text for details.}
\label{rad.ps}
\end{figure}

\begin{figure}
\caption{Luminosity (in \lo) -- mass (in \mo) isochrones at $10^{10}$ years for various metallicities and opacity
models.  Also shown is the zero-metallicity isochrone from Burrows \etal (1993) (``Z models'').
Low-metallicity models drop more precipitously from top right (stars) to bottom left (brown dwarfs).
Furthermore, the lower the metallicity the brighter the star, but the dimmer the brown dwarf.
Measurements for the CM Draconis eclipsing binary system are superposed, and fit the Allard and Hauschildt (1995)
solar-metallicity models nicely. (The boundary conditions for many of these models were provided
by Didier Saumon.)}
\label{LM.metal}
\end{figure}

\begin{figure}
\caption{The same as Fig. \ref{LM.metal}, but for \teff (in K) versus mass isochrones at $10^{10}$ years.
Low-metallicity stars are hotter, while low-metallicity brown dwarfs are cooler. See text for a discussion.}
\label{TM.metal}
\end{figure}

\begin{figure}
\caption{The evolution due to thermonuclear burning of the deuterium mass fraction for various SMO models for
the same mass set as listed in Fig. \ref{lum.ps}.  The initial deuterium mass fraction assumed was 
$2\times 10^{-5}$.  As indicated in the figure, 
the 13 \mj model is the lowest-mass model to show appreciable deuterium
burning.  Moreover, the more massive models burn deuterium more quickly and more completely.
See \S\ref{over} and \S\ref{burn} for details. }
\label{yd}
\end{figure}

\begin{figure}
\caption{The same as Fig. \ref{yd}, but for lithium burning. The fraction of the initial
abundance at a given age and mass is plotted.  The color scheme is as in Fig. \ref{lum.ps}. 
(No SMOs with masses below 13 \mj participate in any lithium burning and so there are no red lines.)
The 0.06 \mo (63 \mj) model is the lowest mass model to
burn appreciable stores of lithium.  See \S\ref{over} and \S\ref{burn} for further details.
}
\label{yl}
\end{figure}

\newpage

\begin{figure}
\caption{This figure depicts the evolution of \teff (in K) with age for the mass set used in Fig. \ref{lum.ps}
and with the same color scheme.  Superposed are dots which mark the ages for a given mass at which 50\% of the deuterium (gold)
and lithium (magenta) are burned. Though the L and T dwarf regions are as yet poorly determined and are no doubt
functions not only of \teff, but of gravity and composition, approximate realms for the L and T dwarfs are indicated
with the dashed horizontal lines.  The spectral type M borders spectral type L on the high-temperature side.
Note that the edge of the hydrogen-burning main sequence 
is an L dwarf and that almost all brown dwarfs evolve
from M to L to T spectral types. 
}
\label{teff}
\end{figure}

\begin{figure}
\caption{Iso-\teff lines in mass (in units of \mj) and age (in years) space for solar-metallicity SMO models.
The colors have no significance, other than to discriminate \teffs. If the effective temperature is known,
its possible mass/age trajectory is given by this plot.  Furthermore, if the age is also known, 
the SMO's mass can be read off the figure.  
}
\label{consttemp.ps}
\end{figure}

\begin{figure}
\caption{A plot of the abundance of the elements versus atomic number.
The position of the element name indicates its elemental abundance according to
Anders and Grevesse (1989) (see Table \ref{anders}).  
The balloons contain representative associated molecules/atoms/condensates of importance
in brown dwarf and EGP atmospheres.  See \S\ref{chem} in text for discussion.
}
\label{elem:abund}
\end{figure}

\begin{figure}
\caption{Depicted are the abundances of the major O, C, and N compounds versus layer temperature
for a representative Gliese 229B model.  N$_2$ and NH$_3$ are in green, CO and CH$_4$ are in red, and 
water is in blue.  The transition from N$_2$ to NH$_3$ occurs near $\sim$700 K, while that
from CO to CH$_4$ occurs from near 1200 to near 1800 K, depending upon the actual pressure/temperature profile.
This model is for \teff = 950 K and a gravity of $10^5$ cm s$^{-2}$.  See \S\ref{chem} for details. 
}
\label{comp.gl229b}
\end{figure}


\begin{figure}
\caption{
The fractional abundances of different chemical species involving
the alkali elements Li, Na, K and Cs for a Gliese 229B model, with rainout
as described in Burrows and Sharp (1999).
The temperature/pressure profile for a \teff=950 K and
$g=10^5$ cm s$^{-2}$ model, taken from Burrows \etal (1997), was used.
Each curve shows the fraction of the alkali element in the indicated form
out of all species containing that element.
All species are in the gas phase except for the condensates, which are in braces \{
and \}.  The solid curves indicate the monatomic gaseous species Li, Na,
K and Cs, the dashed curves indicate the chlorides,
the dot-dashed curves indicate the hydrides and the triple dot-dashed curve
indicates LiOH.
Due to rainout, at lower temperatures there is a dramatic difference
from the no--rainout, complete equilibrium calculation (Fig. \ref{norain}); high albite
and sanidine do not appear, but instead at a much lower temperature the
condensate Na$_2$S (disodium monosulfide) forms, as indicated by the solid
line in the lower left of the figure.  The potassium equivalent, K$_2$S, also
forms, but it does so below 1000 K.  However, K at low temperatures is probably mostly in the form of KCl(s).
The difference between this figure and Fig. \ref{norain} is that almost all the silicon
and aluminum have been rained out at higher temperatures, so that no
high albite and sanidine form at lower temperatures.
}
\label{rainout}
\end{figure}

\begin{figure}
\caption{
The fractional abundances of different chemical species involving
the alkali elements Li, Na, K and Cs for a Gliese 229B model, assuming complete (true)
chemical equlibrium and no rainout (disfavored).  The temperature/pressure profile for a \teff=950 K and
$g=10^5$ cm s$^{-2}$ model, taken from Burrows \etal (1997), was used.
Each curve shows the fraction of the alkali element in the indicated form
out of all species containing that element, {\it e.g.}, in the case of sodium, the
curves labeled as Na, NaCl, NaH and NaAlSi$_3$O$_8$ are the fractions of that
element in the form of the monatomic gas and three of its compounds.  All
species are in the gas phase except for the condensates, which are in braces \{
and \}.  The solid curves indicate the monatomic gaseous species Li, Na,
K and Cs and the two condensates NaAlSi$_3$O$_8$ and KAlSi$_3$O$_8$, {\it i.e.}, high
albite and sanidine, respectively, the dashed curves indicate the chlorides,
the dot-dashed curves indicate the hydrides and the triple dot-dashed curve
indicates LiOH.}
\label{norain}
\end{figure}

\begin{figure}
\caption{The logarithm (base ten) of the pressure (in atmospheres) versus the temperature (in K)
for various brown dwarf models at 1 Gyr (in black, taken from Burrows \etal 1997) and for Jupiter.
The yellow dots denote the positions of the photospheres, defined as where T = \teff. 
Superposed on this figure are various condensation and composition transition lines, 
as well as cloud graphics indicating the approximate position of a cloud base.  
The green lines depict
the T/P trajectory for which the abundance of a neutral alkali atom equals that of its chloride,
ignoring rainout (!).  Also shown in blue are the enstatite, forsterite, and spinel condensation lines.
In red on the left are the ammonia and water condensation lines and on the right are the iron and 
perovskite condensation lines.  In dashed red are the CO/CH$_4$ = 1 and NH$_3$/N$_2$ = 1 lines.
Inner adiabats for 0.08 \mo and 0.09 \mo models are also shown.  
Note that the cool upper reaches of an atmosphere are in the upper left.
}
\label{profile:cloud}
\end{figure}

\begin{figure}
\caption{The absorption cross sections per molecule (in cm$^2$) versus wavelength from the optical to the $M$ band
for water (blue), methane (red), ammonia (gold), molecular hydrogen (green), and
carbon monoxide (purple) at a temperature of 2000 K and a pressure of 10 bars.  Also shown
are the positions of the $Z$, $J$, $H$, $K$, and $M$ bands.
}
\label{opac1500}
\end{figure}

\begin{figure}
\caption{Plotted is the abundance-weighted cross-section spectrum for the
neutral alkali metals Na, K, Cs, Rb, and Li at 1500 K and 1 bar
pressure, using the theory of BMS.  The most important spectral lines for each species
are clearly in evidence.}
\label{opac.nak.wei}
\end{figure}

\begin{figure}
\caption{ A comparison of a recently-generated (unpublished) low-resolution spectral model at \teff{=950} K
and $g$=10$^5$ cm s$^{-2}$ (in red) with the Leggett \etal (1999) spectrum of Gliese 229B (in gold).  The characteristic spikes
at $Z$, $J$, $H$, and $K$ are due to flux streaming through the holes in the water absorption spectrum.  The $\sim$1.7 \mic
and 3.3 \mic bands of methane are readily apparent. This is a generic T dwarf spectrum in the near infrared.
}
\label{gl229b.fit.5}
\end{figure}

\begin{figure}
\caption{The flux (in microJanskys) versus wavelength (in microns) from 1.0 \mic to 10 \mic for a Jupiter-mass
EGP in isolation at various epochs (0.1, 0.5, 1.0, and 5.0 Gyr) during its evolution.  These spectra are compared with the
corresponding black bodies (dashed blue), with \teffs of 290 K, 190 K, 160 K, and 103 K.  Superposed are the approximate sensitivities 
of NICMOS (black dots; Thompson 1992), SIRTF (olive lines; Erickson 1992), and 
Gemini/SOFIA (light green/solid blue; Mountain, Kurz, and Oschman 1994).  A distance of
10 parsecs is assumed.  The positions of the $J$, $H$, $K$, and $M$ bands are indicated at the top.
For all epochs, the super-black-body excess at suitably short wavelengths is always large.
} 
\label{spec.1mj}
\end{figure}

\begin{figure}
\caption{The composition-weighted sum of the absorption cross sections in an SMO atmosphere 
(in cm$^2$) versus wavelength (in microns) from 0.5 \mic to 5.0 \mic.  The blue curve is
at a temperature of 1000 K and a pressure of 1 bar and the red curve is at a temperature of
2200 K and a pressure of 1 bar.  At the higher temperature, TiO and VO are still abundant and
dominate the opacity in the optical and near infrared.  At 1000K, TiO and VO have disappeared,
there should be few or no grains, and the neutral alkali metal atoms, Na and K, 
are assumed (for the purposes of this plot) to be still in evidence.
At the longer wavelengths, water dominates at both temperatures.  The differences
between the two curves encapsulate the essential differences between the spectra of the M and T dwarfs.  
}
\label{opac.sum}
\end{figure}

\begin{figure}
\caption{Spectrophotometry with LRIS on Keck~II (Oke \etal 1995; Kirkpatrick \etal 1999) of
representative red spectra spanning the range from the latest M dwarf to the
latest L dwarf subtypes.  Individual objects from top to bottom are
2MASS~J1239+2029, 2MASS~J1146+2230, DENIS-P~J1228.2-1547, and
2MASS~J1632+1904. Some of the relevant molecular and atomic features are indicated.
}
\label{liebert.A}
\end{figure}

\begin{figure}
\caption{The Keck II spectrum of the L5 dwarf, 2MASSW J1507, from $\sim$4000\AA\ to $\sim$10000\AA,
taken from Reid \etal (2000a).   Clearly seen are the K I absorption feature(s) at $\sim$7700\AA,
the strong absorption feature in the Na D line(s), the Cs and Rb lines, and various FeH and CrH bands,
all indicated on the figure (kindly provided by J.D. Kirkpatrick).  An optical color program reveals
that this L dwarf is magenta.  See \S\ref{shape} and \S\ref{spectra} for discussions.
}
\label{1507}
\end{figure}

\begin{figure}
\caption{The Logarithm$_{10}$ of the cross section spectra of the
neutral alkali metals Na(red), K(green), Cs(gold), Rb(blue), and Li(magenta) at 1500 K and 1 bar
pressure, using the theory of BMS for the K and Na resonance features.
The subordinate lines excited at this temperature are included and the cross sections
are not weighted by abundance.  The importance of this plot is
its implicit line list for the neutral alkali metals that figure so prominantly
in L and T dwarfs.
}
\label{opac.nak.ind}
\end{figure}

\begin{figure}
\caption{The transition from L to T in near-infrared spectra courtesy
of S. Leggett and Reid (2000). Data for Gl 229B are taken from Geballe \etal 
(1996). Superposed are the nominal response curves
for the $J$ (1.2$\mu$m), $H$ (1.6$\mu$m) and $K$ (2.2$\mu$m) photometric bands. 
}
\label{liebert.C}
\end{figure}

\begin{figure}
\caption{Absolute $J$ versus $J-K$ color--magnitude diagram.  Theoretical isochrones
are shown for $t$ = 0.5, 1, and 5 Gyr, along with their blackbody
counterparts.  The difference between blackbody colors and model
colors is striking.  The brown dwarf, Gliese 229B,
the L dwarfs Calar 3, Teide 1, and GD 165B,
and the very late M dwarf LHS 2924 
are plotted for comparison (Oppenheimer \etal 1995; Zapatero-Osorio, Rebolo, and Martin 1997;
Kirkpatrick, Henry, and Simons 1994,1995).
The lower main sequence is defined by a selection of M--dwarf stars from
Leggett (1992). Figure taken from Burrows \etal (1997).
}
\label{JJK}
\end{figure}

\begin{figure}
\caption{The transition from L to T in near-infrared colors (from
Reid 2000). Red crosses are nearby stars, solid purple points are L dwarfs, green 
triangles are Gl 229B-like T dwarfs, and yellow squares are early T dwarfs from
Leggett \etal (2000b).
}
\label{liebert.D}
\end{figure}

\begin{figure}
\caption{The M$_J$ versus $J-K$ near-infrared color-magnitude diagram (from Reid 2000).
Open triangles identify stars from the 8-parsec sample with photometry by
Leggett (1992), red crosses mark nearby stars with 2MASS $JHK_s$ data, solid purple points
are late-M and L dwarfs with photometry from either 2MASS or USNO, and aqua 
points in the bottom left are the T dwarfs Gl 229B (far left) and Gl 570D.
The spectroscopic subtype (K7 -- L8) associated with the objects shown is given 
on the diagram in blue.
} 
\label{liebert.E}
\end{figure}

\begin{figure}
\caption{
The log of the absolute flux (F$_\nu$) in milliJanskys versus wavelength
($\lambda$) in microns from 0.5 \mic to 1.45 \mic for Gliese 229B,
according to Leggett \etal (1999) (heavy solid), and for four theoretical models (light solid) described in BMS.  Also
included is a model, denoted ``Clear" (dotted),  without alkali metals and without any ad hoc absorber
due to grains or haze.  The horizontal bars near 0.7 \mic and 0.8 \mic denote the
WFPC2 $R$ and $I$ band measurements of Golimowski \etal (1998).}
\label{gl229b.K}
\end{figure}

\begin{figure}
\caption{The ``brightness'' temperature (in K) versus wavelength (in microns) from 0.5 \mic to
5.0 \mic for a representative model of Gliese 229B's spectrum.  This is 
the temperature in the atmosphere at which the zenith
optical depth at the given wavelength is 2/3.   Shown are the wavelength positions of various 
important molecular and atomic absorption features.  Such a plot crudely indicates the depth
to which one is probing when looking in a particular wavelength bin.  Note the many H$_2$O and
CH$_4$ absorption bands and the Na and K resonance features in the optical. 
}
\label{bright}
\end{figure}

\begin{figure}
\caption{Slant optical depth, i.e., integral of opacity along a
straight path through the atmosphere of planet HD209458b, where
the path is tangent to a sphere with radius $r$ from the planet's
center.  The total optical depth for the path will be the sum of
cloud (orange curve), Rayleigh (blue curve), and molecular opacity
(red dashed curves) optical depths.  Cloud optical depth is largely
wavelength-independent; Rayleigh optical depths and molecular opacity
optical depths are shown for 14 different wavelengths within the
HST/STIS experiment's passband (Brown \etal 2000).  Evidently,
molecular opacity optical depths are strongly wavelength dependent,
and dominate in defining the planet's limb at these wavelengths.
A possible enstatite cloud near 1 bar pressure is too deep to play
a role.  At pressures above 1 bar, refraction can significantly
deflect the path from a straight line, but this effect also occurs
too deep in the atmosphere to be significant.
}
\label{RMP_H3.ps}
\end{figure}

\begin{figure}
\caption{Radius ($R_{\rm p}$) versus mass ($M_{\rm p}$) for (top to bottom):
1) fully adiabatic gas giants with surface temperature determined by radiative
equilibrium with 51 Peg A; 2) gas giants with radiative regions near the surface at the age of 51 Peg A (realistic
gas-giant model);
3) pure-helium giants with radiative/convective structure at the same age; 4) pure H$_2$O
models at zero temperature; 5) pure forsterite (Mg$_2$SiO$_4$) models at zero temperature.
The structures of the H$_2$O and rock planets were determined using
the ANEOS equation of state (Thompson 1990).  Figure taken from Guillot \etal (1996).
}
\label{guil.rock}
\end{figure}

\begin{figure}
\caption{
Theoretical evolution of the radii of HD209458b and $\tau$ Boo b (in \rj; $\sim 7\times 10^4$ km) with age (in Gyrs).
Model I (solid) is for a 0.69 \mj object in isolation (Burrows \etal 1997).  Models A ($A_B = 0.0$;
\teff $\sim$1600 K) and B ($A_B = 0.5$; \teff $\sim$1200 K)) are for a close-in, irradiated HD209458b
at its current orbital distance from birth, using the opacities of Alexander and Ferguson (1994).
Models C ($A_B = 0.0$; \teff$\sim$1750 K; \mp = 7 \mj) and D ($A_B = 0.5$; \teff$\sim$1350 K; \mp = 10 \mj)
are for a close-in, irradiated $\tau$ Boo b, using a similar opacity set.  The formalism
of Guillot \etal (1996) was employed for models A-D.  The ranges spanned by models A and B and by models C and D
for HD209458b and $\tau$ Boo b, respectively, reflect the current ambiguities in the observations
and in the theoretical predictions due to cloud, opacity, composition, and depth of flux penetration uncertainties.
Superposed are an error box for HD209458b (far right) using the data of Mazeh \etal (2000) and one for
$\tau$ Boo b using the data of Cameron \etal (1999) and Fuhrmann \etal (1998). Figure taken from Burrows \etal (2000).
}
\label{transit.rad}
\end{figure}

\begin{figure}
\caption{Logarithm of the fluxes of 51 Peg A and b, according to Seager, Whitney, and Sasselov (2000).  
The upper curve is the stellar flux and the lower curve is the full-face planetary
reflection spectrum.  Assumed present is a homogeneous enstatite cloud consisting
of particles with a mean radius of 0.01 \mic.  The planetary reflection
spectrum clearly shows absorption due to the Na D lines, the K I resonance
lines at 7700 \AA, water, and methane (in particular at 3.3 \mic).
}
\label{seager}
\end{figure}

\begin{figure}
\caption{Full-face reflection spectra of representative EGPs with
a range of theoretical equilibrium \teffs.  The basic approach of 
Sudarsky, Burrows, and Pinto (2000, SBP) was employed.  Shown are 
HD209458b (Class V?), 51 Peg b (Class IV ?), Gl 86Ab (Class III), and
55 Cnc c (Class II).  The alkali metal, water, and methane features 
are most prominent and cloud models as described in SBP were used.
See \S\ref{albedo} in text for discussion.
}
\label{refspect}
\end{figure}

\begin{figure}
\caption{
This figure shows, in a similar format to
Fig. \ref{profile:cloud} (which compare), but at much higher pressures
and temperatures, the interior P-T profiles for Jupiter, Saturn,
and a representative brown dwarf (Gl 229B). See \S\ref{jovian} in text for discussion.
(The logarithms are base ten.)
}
\label{TPbill}
\end{figure}

\begin{figure}
\caption{
Evolution of the effective temperature with
time for Jupiter and Saturn.  Horizontal lines show
observed values of \teff at the present epoch
(4.57 Gyr). See \S\ref{jovian} in text for discussion.
}
\label{pfaff}
\end{figure}

\begin{figure}
\caption{The theoretical flux (in microJanskys) versus wavelength (in microns) from 1.0 \mic to 30 \mic for a 15 \mj 
brown dwarf in isolation at various epochs ($10^7$, $10^8$, and $10^9$ years) during its evolution.  
The corresponding \teffs are 2225 K, 1437 K, and 593 K.
Superposed are the putative sensitivities
of SIRTF (red dashed), NGST (solid red), TPF (solid black), and Keck (dashed black).  A distance of 
10 parsecs is assumed. 
}
\label{15MJ-30}
\end{figure}

\end{document}